\documentclass[3p]{elsarticle}

\usepackage{bm}
\usepackage{amsmath}
\usepackage{amssymb}
\usepackage{color}
\usepackage{xcolor}
\usepackage{array}
\usepackage{siunitx}
\newcolumntype{M}[1]{>{\centering\arraybackslash}m{#1}}
\definecolor{darkgreen}{rgb}{0.1,0.4,0.1}
\definecolor{orange}{rgb}{1,0.65,0.0}

\newcommand{\mb}[1]{\mathbf{#1}}

\journal{Advances in Colloid and Interface Science}

\begin{document}

\begin{frontmatter}

\title{Particle resuspension from complex surfaces: current knowledge and limitations}

\author[oca]{Christophe Henry\corref{cor1}}

\cortext[cor1]{Corresponding author}
\ead{christophe.henry@mines-paris.org}

\address[oca]{Laboratoire Lagrange, Universit\'e C\^ote d'Azur, CNRS, OCA, Bd.\ de l'Observatoire, Nice, France}

\begin{abstract}
 This review explores particle resuspension from surfaces due to fluid flows. The objective of this review is to provide a general framework and terminology for particle resuspension while highlighting the future developments needed to deepen our understanding of these phenomena. For that purpose, the manuscript is organized with respect to three mechanisms identified in particle resuspension, namely: the incipient motion of particles (i.e. how particles are set in motion), their migration on the surface (i.e. rolling or sliding motion) and their re-entrainment in the flow (i.e. their motion in the near-wall region after detachment). Recent measurements and simulations of particle resuspension are used to underline our current understanding of each mechanism for particle resuspension. These selected examples also highlight the limitations in the present knowledge of particle resuspension, while providing insights into future developments that need to be addressed. In particular, the paper addresses the issue of adhesion forces between complex surfaces - where more detailed characterizations of adhesion force distributions are needed - and the issue of particle sliding/rolling motion on the surface - which can lead to particles halting/being trapped in regions with high adhesion forces.
\end{abstract}


\end{frontmatter}

\tableofcontents

\vspace{15pt} 
\hrule
\vspace{30pt} 

\section{Introduction} 
 \label{sec:intro}

 \subsection{Particle resuspension phenomena}
  \label{sec:intro_resusp}
 
Particle resuspension is concerned with how particles adhering to a surface are detached from a surface and re-entrained into the flow. As emphasized in recent reviews \cite{boor2013monolayer, gradon2009resuspension, henry2014progress, ziskind2006particle}, it has received renewed attention over the recent years due to its key role in a number of applications, including:
\begin{itemize}
 \item environmental issues, with for instance dust resuspension by wind \cite{bagnold2012physics, kok2012physics, de2014effects} or sediment dynamics \cite{coleman2008unifying};
 \item medical applications, where resuspension of hazardous airborne particles can occur through walking-induced resuspension \cite{kubota2013aerodynamic} or due to ventilation systems \cite{kim2010source};
 \item industrial concern, such as with the re-entrainment of radioactive particles in nuclear power plant accidents \cite{stempniewicz2010comparison}.
\end{itemize}

This brief non-exhaustive list shows that resuspension affects a large variety of fields with a large spectrum of spatial or temporal scales. It can indeed involve \textbf{particles with different sizes} such as small colloidal particles (in the micrometer scale such as bacteria), suspended particles (within the millimeter scale as silt or sand) or even large inertial particles (such as stones). Meanwhile, the \textbf{timescales associated to resuspension span several order of magnitude}: for example, in the context of aerosol resuspension, a distinction is made between short-term and long-term resuspension (see for instance \cite{braaten1994wind, hall1988measurements, reeks2001kinetic}). Short-term resuspension is characterized by the detachment of particles that are weakly bounded to the surface and can occur within a couple of seconds, whereas long-term resuspension involves particles that strongly adhere to the surface and can take place at much longer times (up to months or even years in the context of radioactive aerosol particles). Particle resuspension displays \textbf{different behavior depending on the nature and morphology of the deposit formed} \cite{henry2014progress,kok2012physics,theerachaisupakij2003reentrainment}. In particular, a distinction is often made between two states, namely monolayer resuspension - where single particles are resuspended from a surface covered by a small amount of deposits such that interaction between particles is negligible - and multilayer resuspension - where particles form a complex multilayered structure such that inter-particle interactions significantly affect resuspension. The case of multilayer resuspension involves issues that are similar to those in granular media.

These selected examples also highlight that particle resuspension is \textbf{a highly multidisciplinary topic}. It involves indeed notions related to fluid dynamics (since resuspension changes with the flow above the surface), to interface forces (high sticking forces can prevent resuspension), to material physics (roughness on surfaces can significantly change the resuspension rate of particles) or to more specific fields depending on the case considered (biological forces affect the sticking of organisms/cells on a surface while capillary forces occur in humidic environments). 

 \subsection{Terminology}
  \label{sec:intro_term}
  
\begin{figure}
 \centering
 \includegraphics[width=0.85\textwidth]{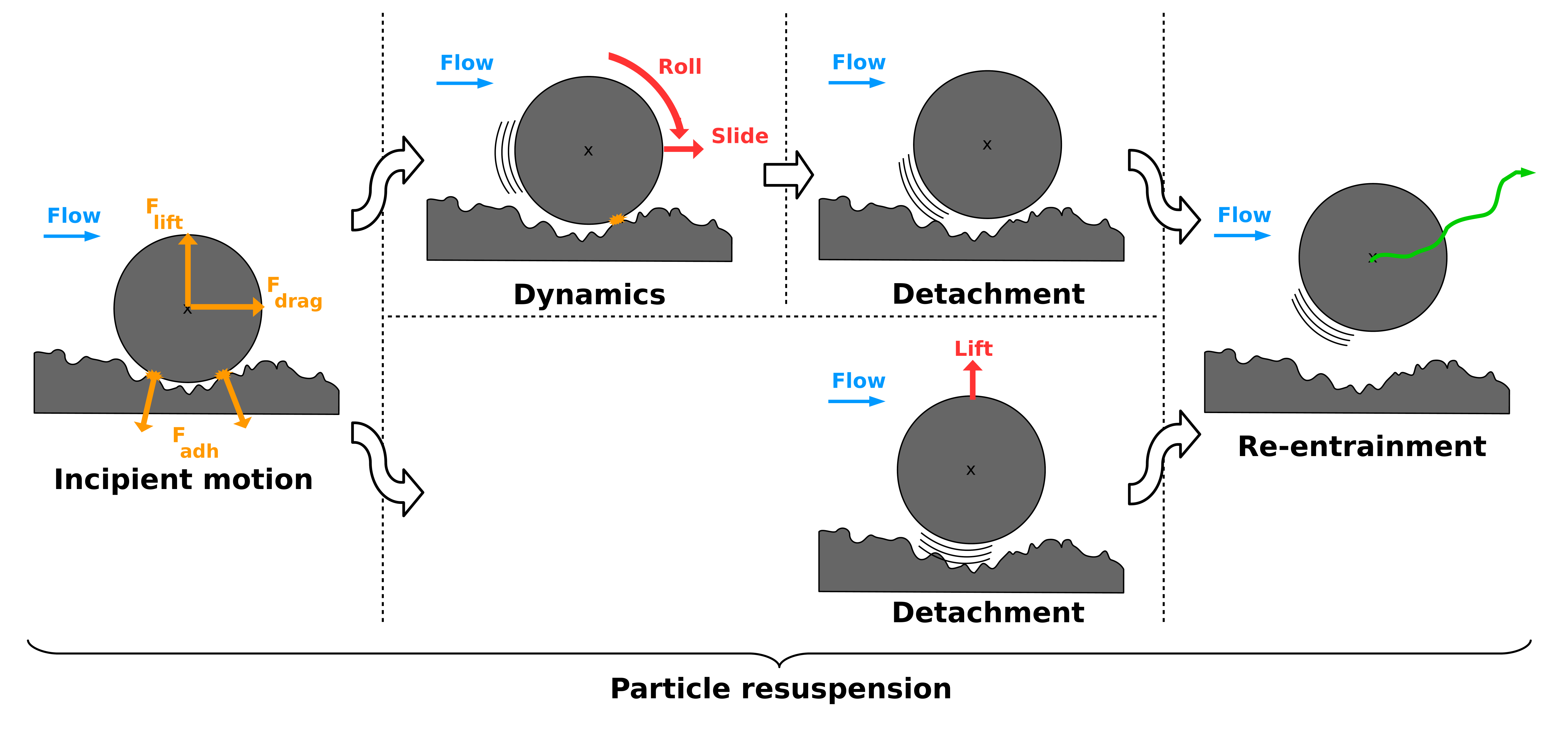}
 \caption{Sketch summarizing the terminology used in the present paper for particle resuspension}
 \label{fig:sketch_def}
\end{figure}

Due to the various fields affected by resuspension, the terminology related to particle resuspension varies. This sometimes leads to confusion regarding the mechanisms at play in resuspension. To avoid such confusion, the key notions that will be used throughout the paper are defined in the following.
\begin{enumerate}
 \item[(def-a)] Sticking/deposited/adhering particles correspond to particles that are in contact with the surface at a given location (no motion).
 \item[(def-b)] Attached particles are particles in contact with the surface but that can be moving (it thus encompasses sticking particles).
 \item[(def-c)] Detached particles refers to the case of attached particles whose contact with the surface has been broken.
 \item[(def-d)] Adhesion forces refer to the contact forces between a particle and a surface.
 \item[(def-e)] Cohesion forces designate the contact forces between particles.
\end{enumerate}
Apart from these fundamental definitions, the following terminology will be used regarding particle resuspension specifically (see also Fig.~\ref{fig:sketch_def}):
\begin{enumerate}
 \item[(def-1)] \textbf{Particle incipient motion} refers to the rupture of the balance between adhesive and moving forces that leads to the motion of sticking particles. The dislodgement of sticking particles triggers their motion either on the surface (attached particles) or in the flow (detached particles).
 \item[(def-2)] \textbf{Particle migration} is related to the motion of attached particles until it is actually \textbf{detached} from the surface.
 \item[(def-3)] \textbf{Particle re-entrainment} refers to the motion of particles after detachment, i.e. in the near-wall region.
 \item[(def-4)] \textbf{Particle resuspension} is a generic term that encompasses all three phenomena (i.e. detachment, migration and re-entrainment).
\end{enumerate}

These definitions allow to distinguish between attached and detached particles while making a clearer distinction between the three mechanisms at play in particle resuspension (i.e. incipient motion, migration and re-entrainment). They remain broad enough to encompass phenomena that have been observed in other fields impacted by resuspension. For instance, in the context of geological flows, abrasion refers to the erosion of a surface by exposure to scraping or other mechanical constraints on the surface: it can thus be interpreted here as a specific case of particle detachment. Another example can be drawn from aeolian transport where different modes for the transport of soil particles (such as sand or dust) have been identified among which saltation and creep \cite{kok2012physics}. Saltation occurs for small materials which display leap-type motion, i.e. they are detached from the surface then carried by the flow before being transported back to the surface. Meanwhile, creep corresponds to the rolling or sliding motion of large particles downstream while staying in contact with the surface. According to the previous definitions, saltation corresponds to a specific situation of particle re-entrainment where sedimentation brings materials back to the surface within a certain finite time. Creep corresponds to a specific situation encountered in particle migration.
 
 \subsection{Purpose of the paper} 
  \label{sec:intro_aim}

With respect to these definitions, the aim of the present paper is to detail our current understanding of the mechanisms related to particle incipient motion, migration and re-entrainment. Since recent reviews have thoroughly described existing models and experimental data (see \cite{henry2014progress, minier2016particles} and references therein), this paper is focused on the physical and theoretical aspects of particle resuspension. Through a set of well-chosen examples, the main limitations of our current understanding of particle resuspension are highlighted. A careful analysis of these examples also provides insights into future experimental/numerical developments that are needed to deepen our understanding of the resuspension phenomenon as well as to refine existing models.

For that purpose, the paper is divided in four sections. Section~\ref{sec:force} briefly presents the typical forces acting on a particle sticking to a surface. Then, the rest of the paper is organised with respect to the three main mechanisms for particle resuspension. The incipient motion of particle is first analyzed in Section~\ref{sec:incipient}, where the rupture of balance between adhering and moving forces is examined. Then, the dynamics of particles moving on a surface is presented in Section~\ref{sec:migration} with a specific emphasis on the role of particles rolling/sliding on the surface since such particles can possibly come to rest on the surface. Section~\ref{sec:reentrain} explores the subsequent case of re-entrained particles and their motion in the near-wall region.

\section{Physics of particle suspensions: fundamental equations and forces}
 \label{sec:force}
 
Particles deposited on a surface exposed to a fluid flow are subject to a variety of forces that can act either to keep the particle attached to the surface or to detach it. Adopting a Lagrangian point of view, the governing equations for particle dynamics are obtained by applying the fundamental laws of classical mechanics giving:
\begin{eqnarray}
\frac{d\mb{x}_{\rm p}}{dt} & = & \mb{U}_p, \\
m_{\rm p}\frac{d\mb{U}_{\rm p}}{dt} & = & \mb{F}_{\rm f \to p} + \mb{F}_{\rm s \to p} + \mb{F}_{\rm p \to p} + \mb{F}_{\rm ext}, \\
I_{\rm p}\frac{d\mb{\Omega}_{\rm p}}{dt} & = & \mb{M}_{\rm f \to p} + \mb{M}_{\rm s \to p} + \mb{M}_{\rm p \to p} + \mb{M}_{\rm ext} 
\label{eq:part_transport}
\end{eqnarray}
with $\mb{x}_{\rm p}$ the particle position, $\mb{U}_{\rm p}$ its translational velocity, $\mb{\Omega}_{\rm p}$ its rotational velocity, $m_{\rm p}$ its mass, $I_{\rm p}$ its moment of inertia, $\mb{F}_{\rm f \to p}$ and $\mb{M}_{\rm f \to p}$ the hydrodynamic forces and torques acting on particles, $\mb{F}_{\rm s \to p}$ and $\mb{M}_{\rm s \to p}$ the surface-particle forces and torques, $\mb{F}_{\rm p \to p}$ and $\mb{M}_{\rm p \to p}$ the particle-particle forces and torques, $\mb{F}_{\rm ext}$ and $\mb{M}_{\rm ext}$ the external forces and torques. 

Even though Eq.~(\ref{eq:part_transport}) is simple and exact, the difficulty comes from the fact that the expressions of the various forces acting on particles are not always straightforward or exact. In the following, we briefly summarize the key forces acting on particles attached to a surface exposed to a flow (interested readers can find more details in \cite{boor2013monolayer, gradon2009resuspension, henry2014progress, marshall2014adhesive, minier2016statistical, ziskind2006particle, ziskind1995resuspension}).

 \subsection{Particle-fluid forces}
  \label{sec:force_part_fluid}

Fluid-particle forces usually include contributions from drag, lift, added-mass and Basset forces as well as Brownian motion: 
 \begin{eqnarray}
  \bm F_{\rm f \to p} = \bm F_{\rm drag} + \bm F_{\rm lift} + \bm F_{\rm add-mass} + \bm F_{\rm Basset} + \rm{\rm Brownian}
  \label{eq:Ff-p}
 \end{eqnarray} 
 \begin{itemize}
  \item Drag forces arise from the friction due to the relative velocity between the fluid and particle velocities. These forces are thus proportional to the relative velocity and to the fluid viscosity. For instance, in the case of spherical particles with a size much smaller than the characteristic scales associated to the fluid (point-particle approximation), the drag force can be approximated by a modified Stokes drag force \cite{clift2005bubbles,gatignol1983faxen,marshall2014adhesive}:
  \begin{equation}
   \mb{F}_{\rm drag} = 6 f \pi R_{\rm p}\rho_{\rm f} \nu_{\rm f}\, \left( \mb{U}_{\rm f} - \mb{U}_{\rm p} \right)
   \label{eq:Fdrag}
  \end{equation}
  with $R_p$ the particle radius, $\rho_f$ the density of the fluid, $\nu_f$ its kinematic viscosity and $\mb{U}_f$ its velocity (approximated here by the velocity at the particle position without Faxen's corrections for large particles \cite{clift2005bubbles,gatignol1983faxen,marshall2014adhesive}). $f$ is a correction factor which includes corrections for fluid inertia and for fluid anisotropy. The anisotropy of the fluid flow in the near-wall region gives rise to different components of the drag in the wall-normal and wall-tangential directions and, as a result, to a non-zero torque force \cite{o1968sphere}.
  \item Lift forces correspond to the force exerted by the fluid flow in the direction perpendicular to the oncoming flow direction (contrary to the drag forces that act along the direction of the relative velocity). The lift force depends on the relative velocity between the particle and the fluid (Saffman lift force) and on the particle rotation rate (Magnus force) \cite{marshall2014adhesive}. As for drag forces, several analytical/empirical formulas have been expressed in the literature and there is no consensus on an exact formulation. For instance, empirical formulas have been written for particles in a fully developed turbulent boundary layer \cite{hall1988measurements, mollinger1996measurement}, which reads:
  \begin{eqnarray}
   \bm F_{\rm lift} = \alpha \nu_f^2\rho_{\rm f} \left(Re_{\rm p}\right)^{\beta}
   \label{eq:Flift}
  \end{eqnarray}
  with $\alpha$ and $\beta$ two constants and $Re_{\rm p}$ the particle-based Reynolds number.
  \item Added-mass forces are associated to the additional force that is needed to accelerate/decelerate a body immersed in a dense fluid due to the fact that the body also moves some of the nearby surrounding fluid as it moves through it. The added-mass force is usually expressed with an added-mass coefficient $C_A$ (taken simply as $C_A = 1/2$) and is a function of the difference between the acceleration of the fluid and the particle acceleration:
  \begin{equation}
   \bm F_{\rm add-mass} = \frac{4\pi R_{\rm p}^3}{3}\, C_A\, \rho_{\rm f} \left( \frac{D\mb{U}_{\rm f}}{Dt} - \frac{d\mb{U}_{\rm p}}{dt}\right)
   \label{eq:Faddmass}
  \end{equation}
  The added-mass force is an unsteady force due to changes in the nearby fluid velocity. It contributes to the dynamics of small particles with a density close to the fluid density $\rho_{\rm p}\sim\rho_{\rm f}$ but it is negligible compared to drag forces for particles much heavier than the fluid $\rho_{\rm p}\gg\rho_{\rm f}$.
  \item Basset history forces are related to the response of the fluid boundary layer to an acceleration of the relative velocity. Due to viscous effects, there is indeed a temporal delay in the development of the boundary layer around a particle when the relative velocity changes, resulting in a 'history' effect. Approximate expressions have been found for a spherical particle in the bulk flow at low Reynolds number \cite{basset1888treatise}:
  \begin{eqnarray}
   \bm F_{\rm Basset} = 6 R_{\rm p} \sqrt{\pi \nu_f \rho_{\rm f}^2 R_{\rm p}^2} \int_{-\infty}^{t} \frac{1}{\sqrt{t-t'}} \, \left(\frac{d\bm U_{\rm f}}{dt'}-\frac{d\bm U_{\rm p}}{dt'}\right)dt'
   \label{eq:Fbasset}
  \end{eqnarray}
  \item Buoyancy forces are associated to the density difference between the fluid and the particle, which gives an acceleration:
  \begin{equation}
   \bm F_{\rm Buoyancy} = \frac{4\pi R_{\rm p}^3}{3}(\rho_{\rm p}-\rho_{\rm f})\bm g
   \label{eq:Fbuoyancy}
  \end{equation}
  Buoyancy forces are thus related to gravity forces: these forces play a role in the dynamics of suspended particles but are negligible compared to Brownian motion for colloidal particles.
  \item Brownian motion corresponds to the random motion of particles suspended in a fluid due to their collision with molecules of the fluid. A usual and straightforward way to include Brownian motion in a particle equation of motion is to add a white noise term. Neglecting other contributions, this gives the following Langevin equation (more details in \cite{henry2014progress,minier2016statistical}):
  \begin{eqnarray}
   d\mb{U}_{\rm p} = K_{\rm Br}d\bm W_i
   \label{eq:Langevin}
  \end{eqnarray}
  where $K_{\rm Br}$ is the diffusion coefficient for Brownian motion and $d\bm W_i$ is the increment of a Wiener process (independent noise terms). Accounting for Brownian motion is relevant when dealing with colloidal particles, i.e. particles that are small enough to be affected by molecular effects (usually $R_p\lesssim\SI{1}{\mu m}$), but is negligible for suspended or larger particles compared to gravity forces.
 \end{itemize}
 
  \subsection{Particle-surface forces}
  \label{sec:force_part_surf}
  
Particle-surface forces are related to the interaction between two bodies in contact. In the case of solid particles attached to a surface, these forces are often dominated by short-range adhesion forces but long-range interactions sometimes need to be taken into account.
 \begin{itemize}
  \item Adhesion of solids arise from short-ranged intermolecular interactions between the molecules composing each solid. These forces can be calculated using various models, which are based on two different methods (see \cite{israelachvili2011intermolecular, henry2014progress, prokopovich2011adhesion} for more details): those based on van der Waals attractive forces \cite{hamaker1937london} and those based on contact mechanics theories (such as JKR \cite{johnson1971surface} or DMT \cite{derjaguin1975effect} theories). The former models describe adhesion as the short-range attractive force between surfaces at very close separation distances due to van der Waals forces (additional terms such as electric double-layer forces, solvation forces, hydrophobic or steric forces can be added to treat systems where such contributions are not negligible). The latter models consider the elastic response of the two bodies in contact, i.e. their deformation, and are based on the surface energy of each body. Both approaches provide an expression for the adhesion force between an spherical particle and a plate that can be written as:
  \begin{eqnarray}
   \bm F_{\rm adh} = g(\rm{model}) \, R_{\rm p} \, \bm e_{\perp}
   \label{eq:Fadh}
  \end{eqnarray}
  with $\bm e_{\perp}$ the wall-normal vector and $g$ a function of the Hamaker constant and separation distance for models based on van der Waals interaction or a function of the surface energies for models based on contact mechanics theories.
  
  Adhesion can be significantly modified in humid environment. In that case, a liquid meniscus can form around the contact area between the particle and the surface due to the condensation of water, resulting in an additional capillary force (as in \cite{ibrahim2004microparticle, zhang2011effects}). Similarly, specific bindings between surfaces can occur when dealing with biological or organic particles such as marine organisms that accumulate on a ship's hull \cite{banerjee2011antifouling}: this includes hydrophobic/electrostatic forces for bacteria adhesion on surfaces or polymer tethering on surfaces to reduce protein adsorption (see also \cite{israelachvili2011intermolecular} for an overview of such specific forces).
  \item Long-range interactions can occur in specific situations. For instance, magnetic interactions between particles and surfaces arise between magnetic materials \cite{ryde2000deposition} while electrostatic interactions can be significant when dealing with charged particles in air \cite{zhang2011effects}.
  \item A friction term results from the contact between a particle and a surface. Dry friction corresponds to the force resisting the relative motion of two solid surfaces in contact. It is thus an energy dissipating force and is part of the science of tribology (interested readers can find more details in \cite{israelachvili2011intermolecular, persson2013physics, persson2013sliding}). This force arise from a combination of various components, including the effects of adhesion forces, surface roughness and surface deformation. It is usually decomposed as a static friction coefficient $\mu_{\rm s}$ (when surfaces have no relative motion) and a dynamic friction coefficient $\mu_{\rm d}$ (when surfaces are in relative motion), which can take different values. This friction coefficient relates the wall-normal component of the forces acting on a particle to a wall-tangential force which acts to prevent the motion of this particle.
 \end{itemize}
 
  \subsection{Particle-particle forces}
  \label{sec:force_part_part}

Particle-particle forces occur when two particles are in proximity of each other. These forces arise when the local concentration of particles is high enough to trigger frequent particle-particle encounters. These interactions thus play a predominant role in dense suspensions, i.e. where the concentration of particles in the solution is high (typically a volume fraction $\Phi\gtrsim10^{-3}$ \cite{elghobashi1994predicting}). Such forces need to be taken into account to properly capture collision, aggregation or fragmentation events. As for particle-surface interactions, these forces can be decomposed as short-ranged and long-ranged interactions. Cohesion forces are short-ranged interactions (the counterpart of adhesion forces for two particles) often modeled using a Lennard-Jones type of potential, which includes van der Waals attractive forces and Pauli repulsion forces (more details can be found in \cite{henry2012towards, liang2007interaction}). Short-ranged electrostatic double-layer interactions are added for charged particles immersed in a liquid while steric forces are needed when dealing with suspensions in polymeric fluids \cite{israelachvili2011intermolecular}. Capillary forces can significantly modify the cohesion forces in a humid environment by creating liquid bridges around the contact area between particles. Long-range interactions can include contributions from electric or magnetic forces in the case of charged/magnetic particles.

 \subsection{External forces}
  \label{sec:force_external}
  
External forces may play a role in specific situations, including for example in the presence of an external electric/magnetic field. These forces are left out of the present paper.

\section{Incipient motion: a balance between adhering and moving forces}
 \label{sec:incipient}

The present section is concerned with presenting how the forces/torques mentioned previously can actually set a particle in motion.

 \subsection{Initial motion: rupture of static equilibrium}
  \label{sec:incipient_initial}
 
\begin{figure}
 \centering
 \includegraphics[scale=0.2, trim = 1cm 1cm 1cm 2cm, clip]{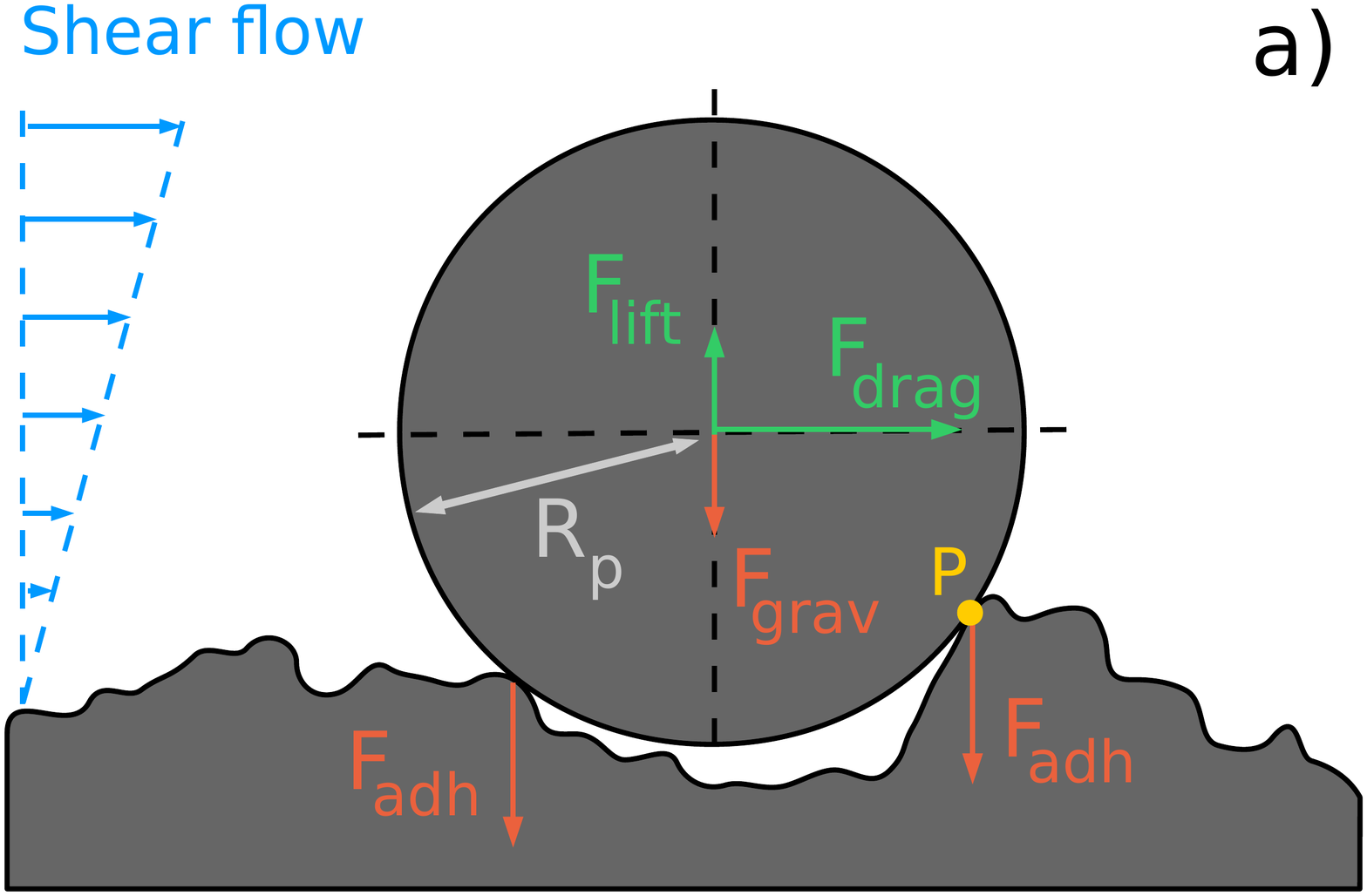} \hspace{50pt}
 \includegraphics[scale=0.2, trim = 1cm 1cm 1cm 1cm, clip]{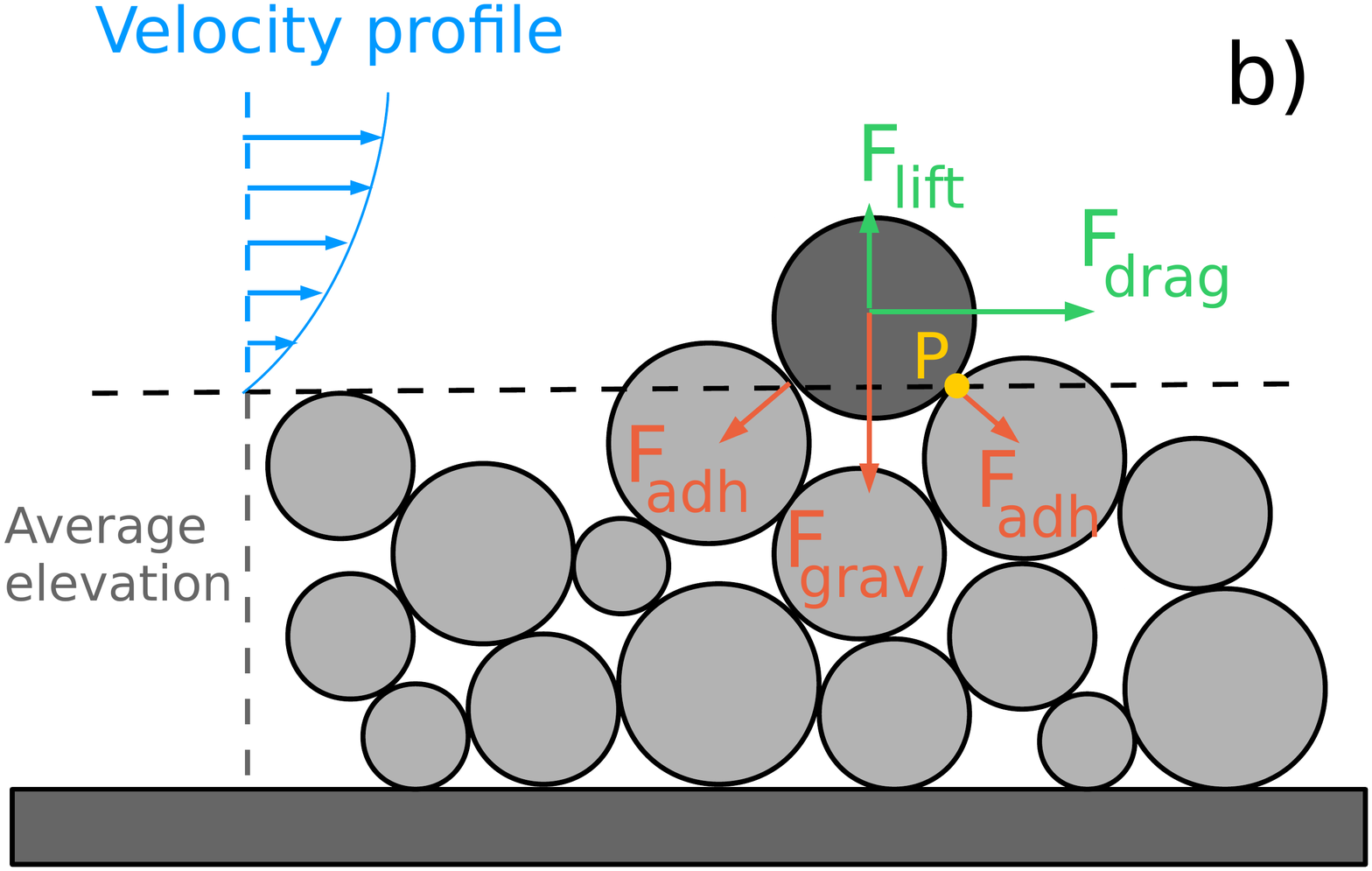}
 \caption{Sketch of a particle sticking to a surface exposed to a flow showing the key forces acting on the particle: a) particle sticking to a clean surface; b) particle sticking on a multilayered deposit.}
 \label{fig:force_part_stick}
\end{figure}

A particle sticking on a surface has no translational/rotational motion, i.e. $\bm U_{\rm p} = 0$ and $\mb \Omega_{\rm p} = 0$, such that Eq.~(\ref{eq:part_transport}) becomes a simple balance equation. This means that adhering particles can start moving when the balance between forces and/or torques is ruptured. It is commonly inferred in the literature that the static equilibrium of a particle sticking on a surface can be broken through three possible mechanisms \cite{wang1990effects}: direct lift-off, rolling or sliding. In the following, the case of heavy solid particles is considered (such that added mass forces are negligible) and only adhesion forces are retained for inter-surface interactions (as depicted in Fig.~\ref{fig:force_part_stick}). Further neglecting Basset forces for simplification, the equations for the incipient motion reduce to:
\begin{eqnarray}
 \rm{Lift-off} & \bm F_{\rm lift} & > \bm F_{\rm grav} + \bm F_{\rm adh}, \\
 \rm{Sliding} & \bm F_{\rm drag} & > \mu_{\rm s} \left(\bm F_{\rm adh} + \bm F_{\rm grav} - \bm F_{lift} \right), \\
 \rm{Rolling} & \bm M_{\rm drag} + \bm M_{\rm lift}& > \bm M_{\rm adh} + \bm M_{\rm grav} 
 \label{eq:incipient}
\end{eqnarray}
where the torques are calculated with respect to the pivot point labeled 'P' in Fig.~\ref{fig:force_part_stick}.

It appears from this brief overview of the forces acting on particles that the relative importance of each force depends on the situation considered (particle size, fluid velocity, nature of surfaces, etc.). For instance, considering the simple case of a particle adhering to a surface exposed to a flow, the incipient motion occurs at lower velocities for large particles since the adhesion torque roughly increases with $R_{\rm p}$ while drag torques are proportional to $R_{\rm p}^2$. Recent measurement of particle resuspension have also shown that Basset history forces can play a role in the case of large particles with a density close to the fluid density \cite{traugott2017experimental}.

\begin{figure}
 \centering
 \includegraphics[width = 0.45\textwidth]{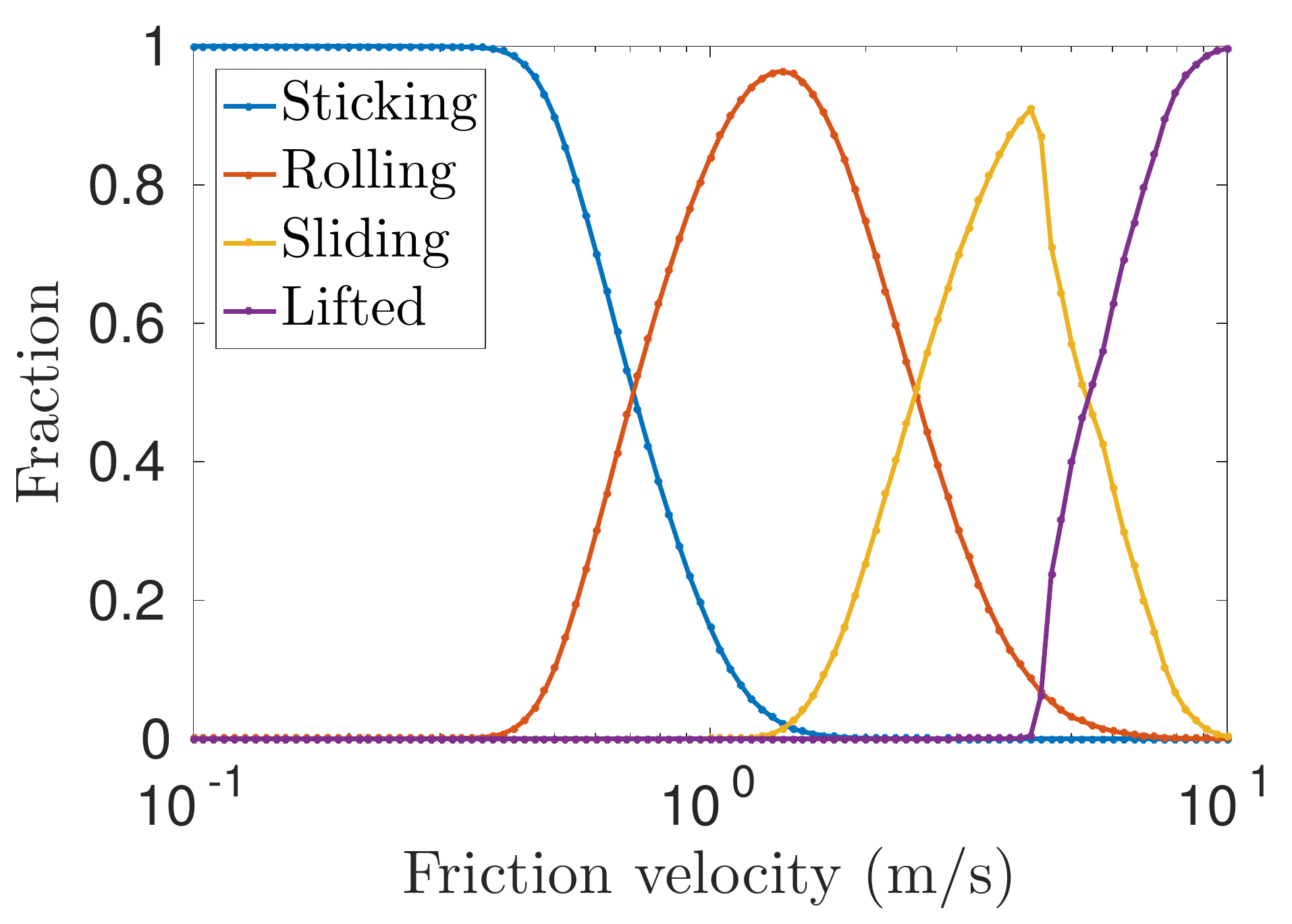}
 \caption{Numerical results showing the relative importance of the various incipient motion (i.e. rolling, sliding, lifting) as a function of the friction velocity. Simulations have been performed using a recent stochastic model \cite{henry2014stochastic} (with extensions for lifting and sliding motion) and considering the case of \SI{5}{\mu m} particles on a rough substrate (\SI{10}{nm} asperities covering \SI{2}{\%} of the surface) exposed to an airflow.}
 \label{fig:incipient_relative_role}
\end{figure}

It is worth noting here that lift-off actually results in immediate particle detachment from the surface whereas rolling and sliding motion give rise to particle migration on the surface: this is also summarised in Fig.~\ref{fig:sketch_def} which underlines these two main roads that lead to particle detachment from a surface. The relative importance of these various mechanisms has been shown to depend on the situation considered \cite{rabinovich2009incipientA, rabinovich2009incipientB}. In particular, it was underlined in a recent review \cite{henry2014progress} that the resuspension of solid spherical particles in a turbulent flow depends on the particle size: small particles fully embedded in the viscous sublayer tend to be rolling on the surface while particles protruding from the viscous sublayer can interact with near-wall coherent structures leading to more frequent direct lift-off. Yet, there is no data showing the relative importance of these mechanisms as a function of the experimental set-up (particle size, fluid velocity, etc.). Such information will be useful to validate existing models for particle resuspension as well as to confirm the range of validity of this scenario for incipient motion. A simulation has been performed using a recent model for particle resuspension \cite{henry2014stochastic} that has been extended to account for all possible incipient motions: numerical results shown in Fig.~\ref{fig:incipient_relative_role} illustrate that particles remain on the surface for very small velocities but start to move as the velocity increases (first rolling then sliding) before lifting motion completely dominate the incipient motion at large velocities. Similar results are expected when the size of the particles is increased at a constant velocity, with the exception that gravity will prevent the motion of very large particles (millimeter-sized or higher depending on the fluid flow).

 \subsection{Direction for future studies}
  \label{sec:incipient_future}
  
In the following, the limitations of our current understanding of incipient motions are presented together with tentative ideas for future experimental/numerical studies that, in the author's opinion, can deepen our current understanding.
  
  \subsubsection{Adhesion forces: the role of complex geometries}
  \label{sec:incipient_future_adhesion}

\begin{figure}
 \includegraphics[width=0.33\textwidth, trim = 0cm 8.9cm 6cm 0cm,clip]{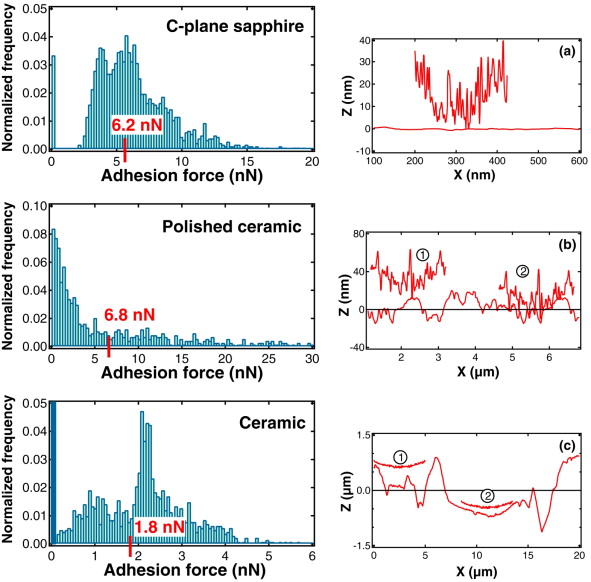}
 \includegraphics[width=0.33\textwidth, trim = 0cm 4.5cm 6cm 4.5cm,clip]{adhesion-distr.jpg}
 \includegraphics[width=0.33\textwidth, trim = 0cm 0cm 6cm 8.9cm,clip]{adhesion-distr.jpg}
 \caption{Distribution of the adhesion force between sapphire spheres and various substrates. The mean value of adhesion is written on the graph. Reprinted from \cite{audry2009adhesion}. Copyright 2009 Elsevier.}
 \label{fig:adhesion_distr}
\end{figure}

Real surfaces differ from ideal smooth surfaces and usually exhibit heterogeneities such as roughness. This results in different torques since the pivot point depends on the locations of particle-surface contacts. Such surface heterogeneities strongly affect adhesion forces and thus incipient motion, as in \cite{shnapp2015comparative}. More precisely, their effect on adhesion forces is two-fold:
\begin{enumerate}
 \item the presence of small-scale features (in the nanometer range) protruding from the surface have been shown to significantly reduce the adhesion force by increasing the separation distance between the core of the two bodies (see for instance \cite{rabinovich2000adhesionI, rabinovich2000adhesionII}). Such lower adhesion forces, which can be several orders of magnitude smaller than the adhesion between smooth surfaces, are responsible for incipient motion at much lower fluid velocities.
 \item the chaotic nature of rough surfaces leads to broad distributions in the adhesion forces. Such variations of the adhesion force is due to the varying number and location of asperities interacting together when rough surfaces are brought in contact and to the different sizes of these surface features. In particular, recent measurements of adhesion forces and resuspension with rough surfaces have shown that the distributions can display different shapes including (see also Fig.~\ref{fig:adhesion_distr}): Gaussian distributions \cite{prokopovich2010multiasperity}, log-normal distributions \cite{gotzinger2004particle}, Weibull distributions \cite{gotzinger2004particle} two-peaks or more complex multiple-peaks distributions \cite{audry2009adhesion, beach2002pull}.
\end{enumerate}

Since distributions in adhesion force have no general forms, it is hard to assume a given form right from the onset. For that reason, there is an increasing literature on experimental measurements of adhesion forces between complex surfaces. Besides specific adhesion models have been recently designed to capture the effect of surface roughness on adhesion forces. Regardless of the physical model retained for the description of adhesion (i.e. based on contact mechanics theories or van der Waals forces), these models account for surface roughness either using very detailed representations (such as Fourier transform or fractal surfaces) \cite{jaiswal2009modeling} or using simplified representations where asperities are randomly placed on a smooth surface \cite{rabinovich2000adhesionII}. Then, the interaction between such rough surfaces is calculated numerically (more details on such models for rough surfaces can be found in \cite{eichenlaub2004roughness, prokopovich2011adhesion} and references therein).

In the context of particle resuspension, two kind of models exist: those based on a given distribution of adhesion force (usually taken from experimental data or assuming a given shape of the distribution) \cite{reeks2001kinetic} and those based on a calculation of this distribution (using adhesion models and measurements of surface roughness characteristics) \cite{guingo2008new, henry2014stochastic}. Both models have been shown to provide accurate predictions of the overall shape of the resuspension rate, especially of $u_{50}^*$ (the friction velocity at which \SI{50}{\%} of the particles are resuspended), provided that the proper forces are accounted for. Yet, these models tend to overestimate the resuspension rate at high friction velocities (see \cite{henry2018colloidal,zhang2013particle}). This discrepancy has been attributed to the tail of adhesion forces which governs the resuspension at high friction velocities since poorly adhering particles are resuspended at smaller friction velocities \cite{henry2018colloidal}.

To go beyond these current limitations, \textbf{more detailed characterizations of surface roughness and its effect on adhesion forces} are needed. For that purpose, new systematic measurements of surface roughness, adhesion forces and incipient motion altogether should be performed with a specific attention on rare events. Recent advances in experimental techniques (using AFM or SEM) have opened new possibilities to obtain precise 2D or 3D images of the surface profile (interested readers can find more details in \cite{whitehouse2004surfaces}). Yet, surface roughness is usually characterized using simple parameters which provide only a limited amount of information (mostly averaged values such as the rms roughness). Besides, this information is extracted assuming spatially homogeneous surfaces and, thus, do not include characteristics on rare local heterogeneities. To obtain more detailed predictions with adhesion models, information on the whole distribution of both the curvature radii (i.e. the local curvature of surface features) and their spatial correlations (i.e. surface coverage) is needed \cite{henry2018colloidal}. In the author's opinion, such detailed information can actually be extracted from surface profile measurements using proper tools (based for instance on a Fourier analysis), as was done recently in \cite{prokopovich2010multiasperity}. Scanning larger portions of the surface will also provide insights into both uniform regions (i.e. statistically homogeneous) and irregular regions (such as isolated large roughness features).

Such refined characterizations will help in designing refined numerical models and will open the way for more systematic studies of long-term resuspension, which is highly affected by rare extreme events that take place on longer times (and usually involve highly energetic events). 

 \subsubsection{Cohesion forces: the role of complex deposits}
  \label{sec:incipient_future_cohesion}
 
Another challenge that needs to be tackled is related to the complex morphologies of the deposit formed by the accumulation of particles on a surface. As depicted in Fig.~\ref{fig:force_part_stick}b, particles deposited on top of the multilayered structures are partially exposed to the fluid flow and these particles are in contact with several other particles forming this bed. The randomness of inter-particle contacts leads to a broad distribution in cohesion forces. In that sense, complex morphologies affect cohesion forces and torques for all range of particle sizes in a similar way to how surface roughness impacts adhesion forces. Besides, this complex morphology also impacts the torques of all forces acting on the particle due to the arbitrary position of the pivot point. The presence of surface roughness on particles can add further complexity (by lowering cohesion forces and changing the positions of contacts) but this will be left out of the present discussion.

\begin{figure}
 \centering
 \includegraphics[scale=0.2]{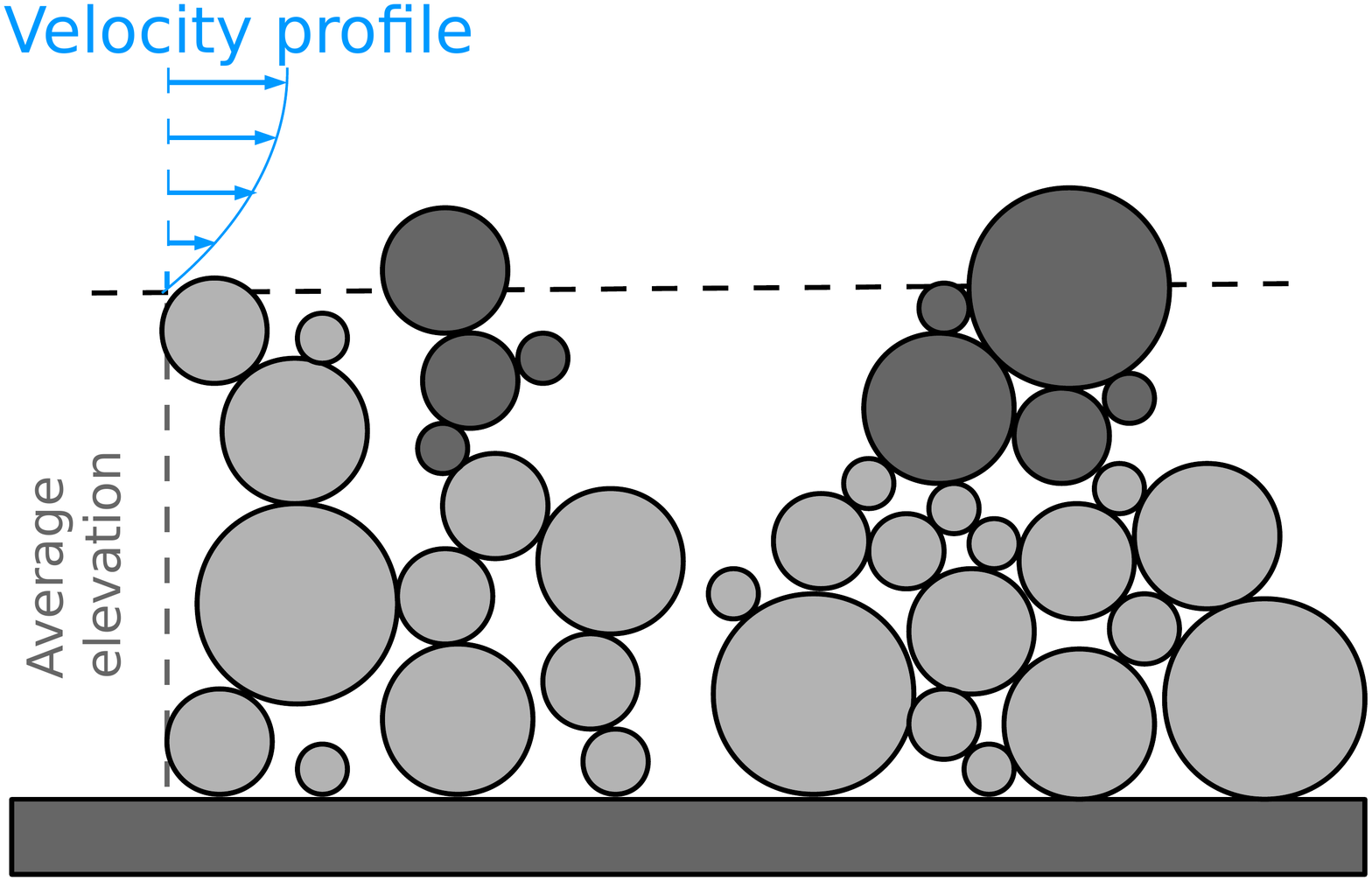}
 \caption{Sketch of a multilayered deposit exposed to a fluid flow: darker particles constitute clusters of particles that can be resuspended.}
 \label{fig:sketch_multilayer}
\end{figure}

To accurately reproduce numerically the incipient motion of such complex deposits, it is necessary to take into account the morphology of the deposit formed (see recent models as in \cite{friess2002modelling, iimura2009simulation}). The key idea of these simulations is to generate a multilayered deposit representative of real deposits and, then, to expose it to a flow. Yet, several difficulties arise when doing such simulations:
\begin{itemize}
 \item the deposit structure is not necessarily fully characterized. The morphology is indeed a direct consequence of how particles accumulate on a surface. In the case of large/heavy particles such as stones, compact deposits are formed since gravity favors the motion of particles until reaching mechanically stable positions, i.e. with isostatic constraints (such as on flat areas or on regions with three contacts). For colloidal particles, cohesion forces dominate over gravity forces and the resulting morphology depends on the deposition scenario \cite{henry2012towards}: loose structures occur when colloids deposit on the multilayered structure at the point of impact whereas more compact structures are obtained when colloids can move until reaching more stable positions.
 \item the structure of the deposit also affects the hydrodynamic forces acting on each particle. Particles inside the multilayered deposit are indeed sheltered by other particles whereas particles protruding from the mean bed elevation are only partially exposed to the fluid flow (see also Fig.~\ref{fig:sketch_multilayer}). This reinforce the anisotropy of the fluid flow around deposited particles and thus changes the corresponding hydrodynamic forces.
 \item incipient motion can involve clusters of particles, especially in the case of polydispersed suspensions as illustrated in Fig.~\ref{fig:sketch_multilayer}. This is due to the lower cohesion forces occurring between small particles than between large ones. As a result, the contact will be broken along regions of low adhesion forces where the force/torque balance is ruptured. This adds further complexity in numerical simulations since incipient motion of a particle requires knowledge of the cohesion forces and torques in nearby regions and not only around the particle. 
 \item cohesion forces are not always constant in time. In the case of solid particles, aging effects such as sintering can lead to an increase in the cohesion forces with time \cite{abd2007influence,francia2015role}. Sintering occurs at high temperature and leads to the creation of stronger bonds between contacting particles due to the precipitation occurring preferentially near the contact area between surfaces. Such modifications of particle bond energies with time are generally referred to as 'consolidation' processes and have been shown to be responsible for lower resuspension rates on heated surfaces after long exposure time \cite{abd2007influence}. When occurring, these consolidation effects thus have a profound impact on long-term resuspension. In the case of alpine snow, higher temperature lead to snow melting and the resulting higher water content decreases the cohesion forces in snow: this process is responsible for wet loose-snow avalanches \cite{mcclung2006avalanche}.
\end{itemize}

Despite these limitations, fine Lagrangian simulations have confirmed the role played by the deposit morphology in multilayer resuspension \cite{iimura2009simulation}. Yet, these simulations were limited to a small number of particles due to the high computational costs associated with the calculation of the cohesion forces in a cluster and of the hydrodynamic forces acting on each particle (which is partially/totally shielded by other particles depending on its position). More systematic studies of the incipient motion in such multilayered systems are still needed experimentally to confirm the impact of the deposit morphology. 

  \subsubsection{Particle shape: effect of non-sphericity}
  \label{sec:incipient_future_nonsphericity}
  
Real particles do not necessarily have a spherical shape and often display complex shapes such as rod-shaped biological agents \cite{brambilla2017adhesion}. In that case, the anisotropy in the particle shape induces hydrodynamic forces that depend on the direction considered and results in complex rotational motions. For that reason, the orientation of non-spherical particles plays a key role in their dynamics. Even though the dynamics and orientation of rigid non-spherical particles has been characterized experimentally and numerically in turbulent flows (more details in \cite{voth2017anisotropic} and references therein), their adhesion and resuspension from surfaces has been seldom studied in the literature \cite{brambilla2017adhesion,stevenson2002incipient}. 

The incipient motion of non-spherical particles on a surface differ from spherical particles due to different hydrodynamic and adhesion forces. In the case of rod-like particles, The expressions for these hydrodynamic and adhesion forces have been discussed in a recent review in the case of rod-like particles \cite{brambilla2017adhesion} while other studies have explored the simplified case of hemispherical particles \cite{stevenson2002incipient}. It was underlined that both adhesion and hydrodynamic forces depend on the orientation of the particle with respect to the underlying surface: particles with their major axis oriented parallel to the surface encounter higher adhesion forces and torques whereas particles with their major axis perpendicular to the surface are subject to higher hydrodynamic forces. This implies that the incipient motion of non-spherical particles depends on both the orientation with respect to the surface and to the direction of the flow.

New experimental and numerical studies are required to investigate what is exactly the effect of non-sphericity on the incipient motion of particles. In the author's opinion, recent advances in image processing techniques \cite{agudo2017detection} can help in designing experimental setup with non-spherical particles. The use of markers on particles coupled with appropriate detection tools provides access to both translational and rotational motion. Such measurements of the initial particle motion can deepen our understanding of the resuspension of complex-shaped particles, especially regarding the frequency of direct lift-off compared to rolling/sliding. This will also provide valuable data for the development and validation of adequate models.

  \subsubsection{Particle shape: effect of flexibility}
  \label{sec:incipient_future_flexibility}
  
\begin{figure}
 \includegraphics[width=0.45\textwidth,trim = 0cm 2.6cm 0cm 0cm,clip]{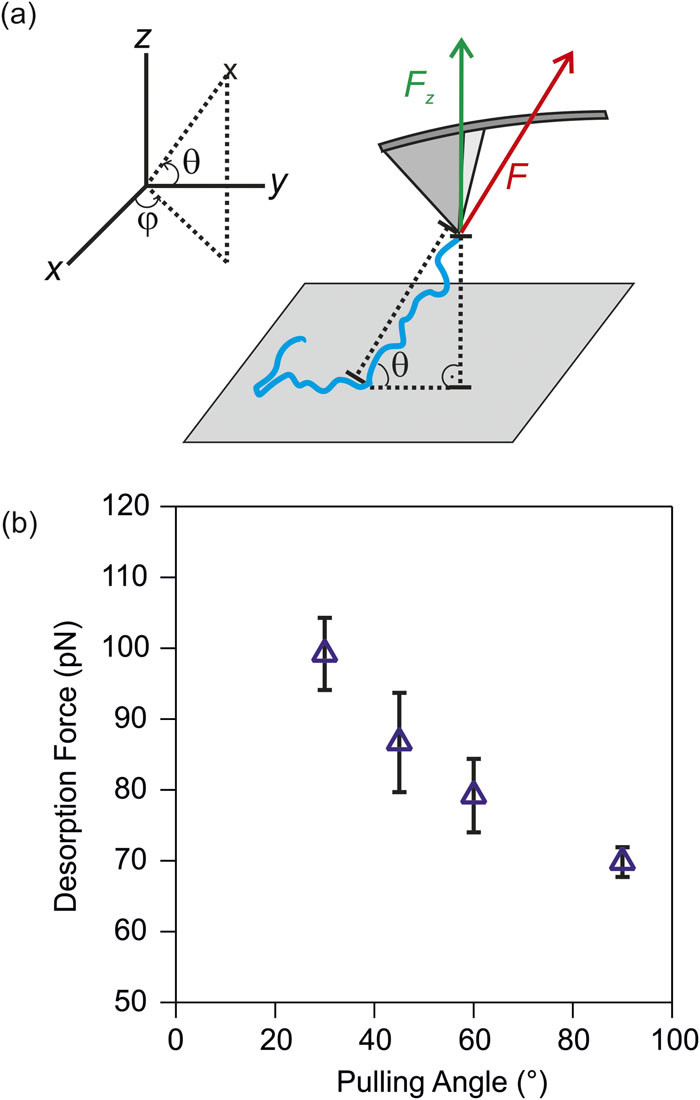}
 \includegraphics[width=0.4\textwidth,trim = 0cm 0cm 0cm 2.0cm,clip]{adhesion-fiber-grebikova}
 \caption{Adhesion of a deformable particle on a surface: (a) Schematic of the experiment, with a polymer partially adsorbed on the surface and pulled by AFM tip; (b) Average desorption force values plotted as a function of the pulling angle. Reprinted from \cite{grebikova2017pulling}. Copyright 2017 with permission from AIP Publishing.}
 \label{fig:adhesion_fiber_grebikova}
\end{figure}

Besides having a complex shape, some particles are flexible: this is the case for instance of fibers \cite{lundell2011fluid}, polymers or DNA \cite{shaqfeh2005dynamics}. The difficulty in dealing with deformable particles lies in the fact that the number of degrees of freedom needed to describe such systems is large. Flexibility thus give rise to a wide range of possible scenarios for the adhesion of deformable particles on a surface: this is illustrated in Fig.~\ref{fig:adhesion_fiber_grebikova}a which displays a polymer in contact with the surface. A polymer can only partially adhere to the surface, i.e. that contact with the surface only occurs on certain portions of the polymer with the remaining parts forming arches above the surface. Therefore, the adhesion force between a polymer and a surface significantly depends on the contact configuration, i.e. the size of the portion actually in contact with the surface and their orientation. Recent advances in experimental techniques now provide access to such measurements by grafting the end of a polymer to an AFM and pulling it away from the surface \cite{grebikova2017pulling}: observations showed that desorption changes according to the pulling angle (see Fig.~\ref{fig:adhesion_fiber_grebikova}b).

In the author's opinion, such procedures should be adapted to the case of larger particles and used to provide systematic and precise data on the adhesion between flexible particles and surfaces. These techniques should also be coupled with proper real-time optical measurements of the particle shape to be able to record the contact configuration between such flexible particles and surfaces. This will indeed deepen our understanding of the effect of flexibility on adhesion. Such data will also help designing fine models to capture such intricate adhesion regimes, using for instance discretization techniques where a flexible particle is represented as a chain of ellipsoids \cite{fan1997flow}.
  
 \section{Particle migration and detachment: dynamics of particles on the surface}
  \label{sec:migration}

  \subsection{Rolling and sliding motion}
   \label{sec:migration_motion}
   
Once the balance between forces/torques is ruptured, a particle is dislodged from its initial sticking position on the surface. Yet, except in the case of lifting motion (where the particle is immediately detached from the surface), the contact with the surface is not broken and the particle starts rolling/sliding on the surface. Particles can then roll/slide for a certain time before being lifted-off the surface as observed sometimes in \cite{kassab2013high}. This means that the incipient motion of a particle is not strictly speaking equal to resuspension (unless all particles are detached by direct lift-off). In the following, we detail the dynamics of attached particles as they move along the surface:
\begin{itemize}
 \item[a.] Rolling motion occur when the torque balance acting on a particle is ruptured. Once rolling motion starts, the dynamics of a rolling particle is simply given by:
 \begin{eqnarray}
 I_{\rm p}\frac{d\mb{\Omega}_{\rm p}}{dt} & = & \mb{M}_{\rm f \to p} + \mb{M}_{\rm s \to p} + \mb{M}_{\rm p \to p} + \mb{M}_{\rm ext} 
 \label{eq:part_rolling}
 \end{eqnarray}
 Such rolling motion has been confirmed by various experimental measurements on small colloidal/suspended particles (see \cite{henry2014progress, kobayakawa2015microscopic} and references therein): it is illustrated in Fig.~\ref{fig:roll_agudo} which shows the motion of a glass bead on a substrate made of smaller spheres \cite{agudo2017detection}.
 In agreement with theoretical expectations from Eq.~(\ref{eq:part_rolling}), Fig.~\ref{fig:roll_agudo} shows that a rolling particle can increase (resp. decrease) its rotational velocity $\mb{\Omega}_{\rm p}$ when torques acting to move the particle exceed (resp. are lower than) torques acting to prevent its motion. As a result, the translational and rotational velocities of rolling particles are not constant and change according to the balance of torques. Furthermore, in the case of rough surfaces, these velocities fluctuate very rapidly since the chaotic nature of roughness leads to rapidly changing forces/torques \cite{henry2014stochastic}. More precisely, considering the case of colloids, the adhesion force on a rough surface can be totally different for a displacement as small as a few percent of the particle diameter.
 \begin{figure}
  \centering
  \includegraphics[width = 0.45\textwidth]{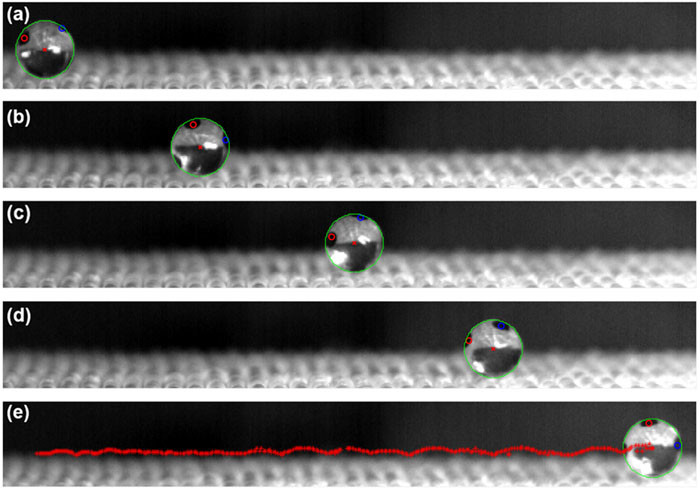}
  \includegraphics[width = 0.45\textwidth, trim = 0cm 0cm 0cm 4.15cm, clip]{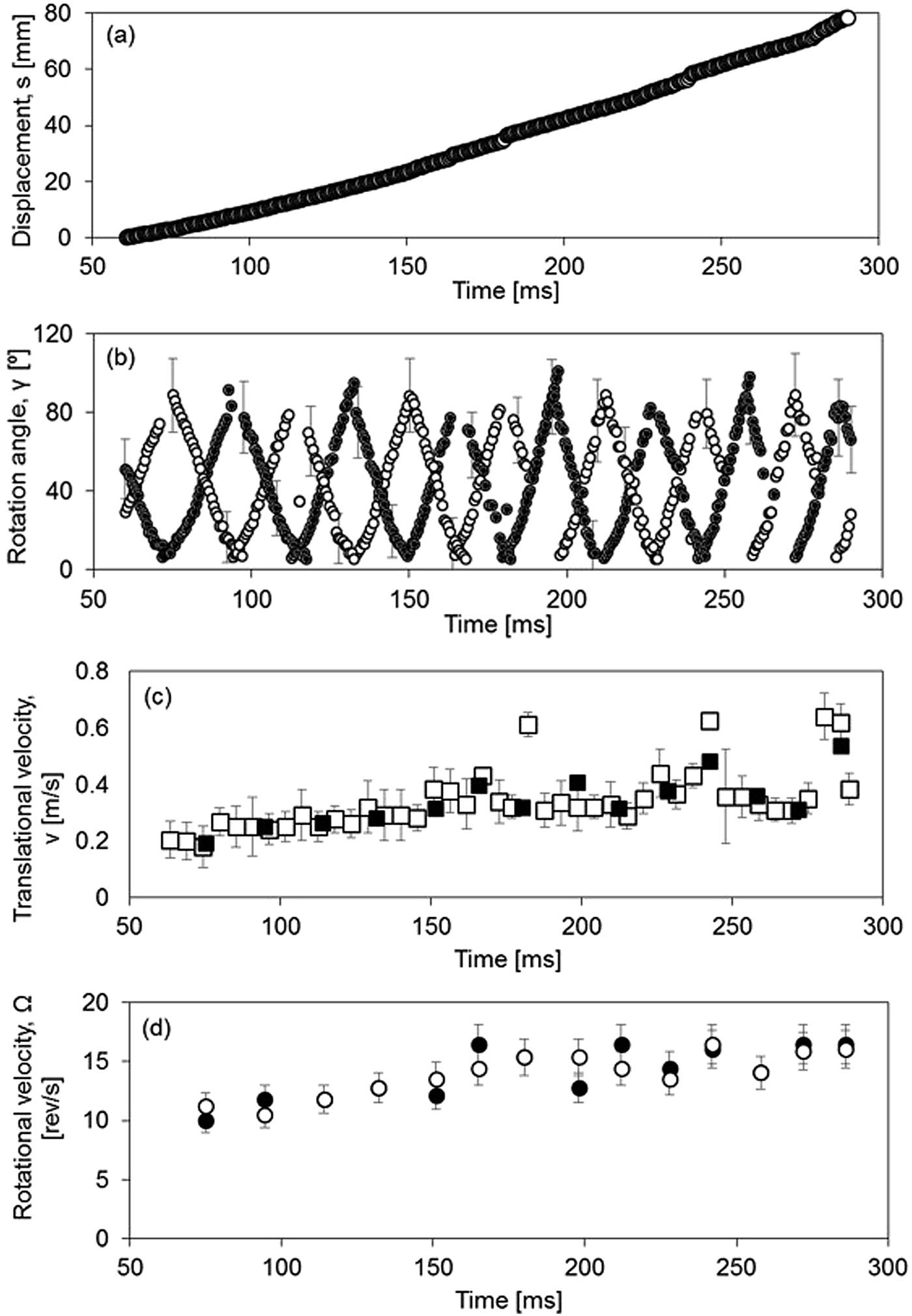}
  \caption{Incipient motion of a glass bead ($7.00\pm 0.35$ mm in diameter) on a substrate made of stationary spheres ($2.00\pm 0.1$ mm in diameter). (Left) Snapshots taken during the motion: red and blued circles indicate the centers of the marks used for image processing and the red crosses in (e) represent the trajectory of the bead center. (Right) Corresponding translational and rotational velocities extracted from these snapshots. Reprinted from \cite{agudo2017detection}. Copyright 2017 with permission from AIP Publishing.}
  \label{fig:roll_agudo}
 \end{figure}
 \item[b.] Sliding motion occur when the force balance acting on a particle is ruptured. Once sliding motion starts, the dynamics of a sliding particle is given by:
 \begin{eqnarray}
 m_{\rm p}\frac{d\mb{U}_{\rm p}}{dt} & = & \mb{F}_{\rm f \to p} + \mb{F}_{\rm s \to p} + \mb{F}_{\rm p \to p} + \mb{F}_{\rm ext}
 \label{eq:part_sliding}
 \end{eqnarray}
 Several experimental studies on colloidal particles have shown that rolling and/or sliding motion can occur (as in \cite{ibrahim2003microparticle, ibrahim2004microparticle}) but these measurements do not allow to distinguish between rolling and sliding. Similarly, large particles (such as sand with a diameter $>\SI{500}{\mu m}$) usually undergo creeping motion, i.e. rolling or sliding motion along the surface \cite{bagnold2012physics, kok2012physics}. The sliding velocity of such particles is also expected to change with the intensity of hydrodynamic/adhesion/gravity forces acting on particles.
\end{itemize}

 \subsection{Detachment from the surface}
  \label{sec:migration_detach}

As a particle moves on a surface, it can be detached from the surface in two ways:

\begin{itemize}
 \item through direct lift-off (hydrodynamic effects). This occurs when the particle interacts with near-wall fluid structures (see \cite{pope2000turbulent, soldati2009physics} and references therein). Turbulent boundary layers are indeed characterized by the existence of near-wall coherent structures, which can be loosely defined as regions of the fluid exhibiting a certain order in space and time. A particle protruding from the viscous sublayer can interact with such high-velocities structures leading to its detachment when the lift force overcomes the forces preventing its detachment (adhesion, etc.). Resuspension caused by such near-wall fluid structures is often referred to as 'burst-type resuspension'. 
 \item through rocking events (mechanical effects). This corresponds to the case when detachment is triggered by an encounter with a specific surface feature (such as a large-scale asperity or another deposited particle) \cite{guingo2008new}. 
\end{itemize}

The intricate coupling between particle migration on the surface and particle detachment from it are relatively new topics and need to be further characterized, especially to assess if the frequency of interactions with such coherent structures changes when particles are moving on the surface.  

 \subsection{Direction for future studies}
  \label{sec:migration_future}
 
  \subsubsection{Rolling/sliding dynamics: role of spatial and temporal scales}

Even though rolling and sliding motion have been confirmed to play a role in particle resuspension, their exact role and importance in resuspension scenarios is not fully understood. This is due to the fact that making a distinction between rolling and sliding motion was not an easy task experimentally while it is actually key in estimating the relative importance of both mechanisms. Such a distinction is now possible using recent image processing techniques coupled with markers on particle surfaces \cite{agudo2017detection} (see also Fig.~\ref{fig:roll_agudo}). For these reasons, experiments need to be conducted using such techniques to extract the proportion of particles undergoing rolling/sliding motion for a wide range of particle sizes. Apart from providing information on the relative importance of both mechanisms for a set of experimental conditions, such measurements will also give access to the major particle characteristics including: their rotational/translational velocity, the time and distance traveled between their incipient motion and their detachment. Some insights of these characteristics have been provided in a recent study of particle motion on an engineered surface (with nanoscale features) \cite{kalasin2015engineering}: using a velocity-based criterion to identify rolling motion, \SI{1}{\mu m} particles were shown to be able to travel distance roughly a hundred times larger than their diameter. The issue is then to see if particles can also travel such large distances on usual real surfaces.

In the author's opinion, these characteristics on the dynamics of particles should be properly captured by refined models due to their potential impact on particle detachment (see previous section). Besides, the distinction between detached particles and particles moving on a surface raises new questions regarding the accuracy of previous measurements of particle resuspension: the number of particles remaining on a certain region of the surface was indeed counted from optical measurements but there is thus no distinction made between detached particles and particles rolling/sliding out of the observation area. This would be especially true if the mean distance traveled by moving particles before actual detachment is large. 
 
  \subsubsection{Irregular motion: potential for arrest/halt}
   \label{sec:migration_halt}

Another aspect related to the dynamics of particles on a surface is their ability to reach regions of the surface where they can stop. From a purely conceptual point of view, this could happen both for monolayer resuspension (high adhesion regions on a surface) and for multilayer resuspension (sheltering in a hole inside a bed arrangement). Recent studies have confirmed that particles can actually stop on a surface when they encounter regions of high adhesion forces/torques \cite{duru2015three, kalasin2015engineering}. Yet, the frequency of occurrence of such events needs to be characterized experimentally and numerically.

\begin{figure}
 \centering
 \includegraphics[width = 0.45\textwidth]{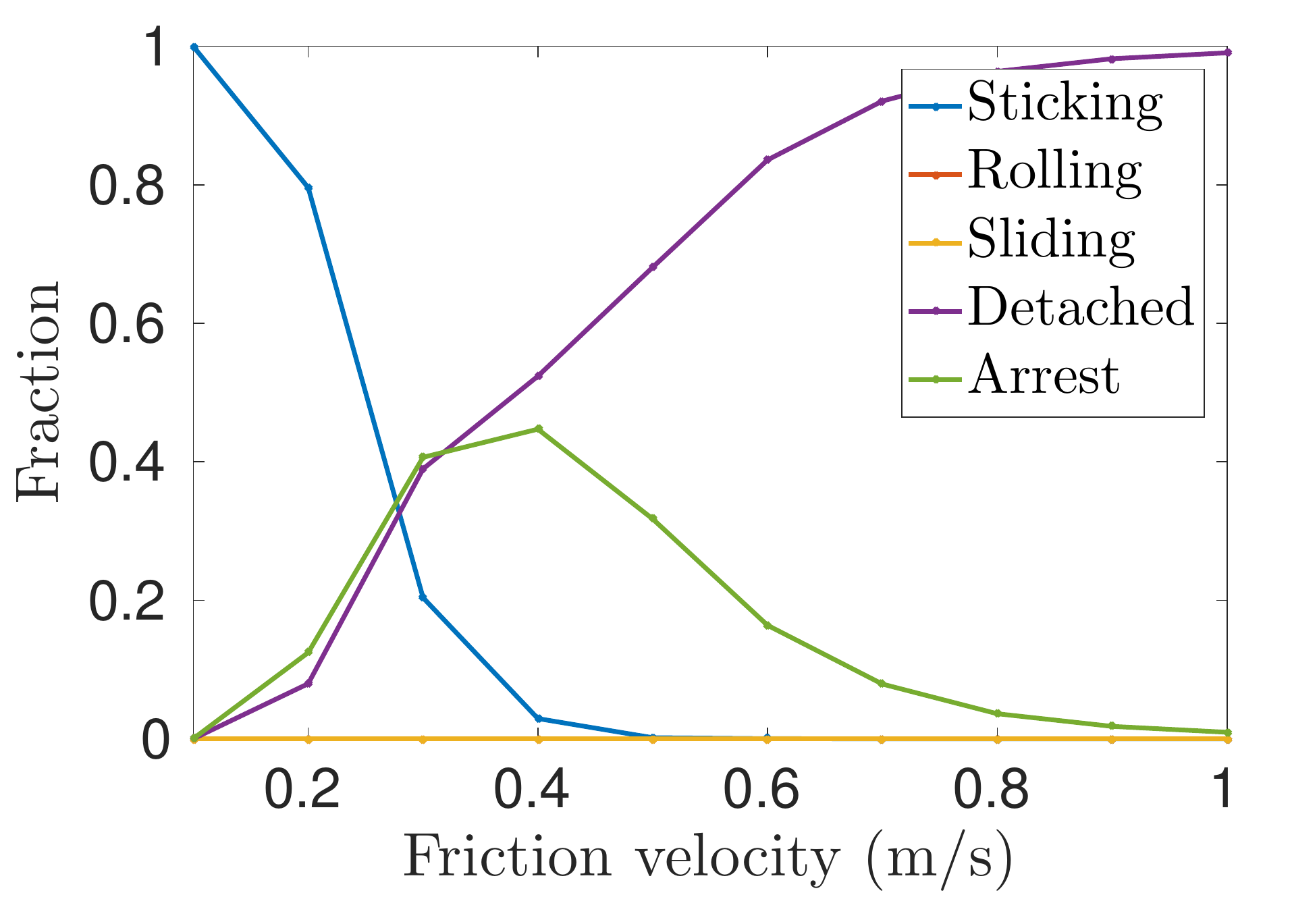}
 \caption{Numerical results showing the relative importance of the migration (i.e. rolling, sliding), detachment and arrest as a function of the friction velocity. Simulations have been performed using a recent stochastic model \cite{henry2014stochastic} (with extensions for lifting and sliding motion) and considering the case of \SI{40}{\mu m} particles on a rough substrate (\SI{25}{nm} asperities covering \SI{2}{\%} of the surface) exposed to an airflow.}
 \label{fig:arrest_relative_role}
\end{figure}

To illustrate this issue, a simulation has been performed using a recent dynamic model for particle resuspension \cite{henry2014stochastic} considering the simple case of spherical glass particles (diameter of \SI{40}{\mu m}) sticking on a rough glass substrate (\SI{25}{nm} asperities covering \SI{1.91}{\%} of the surface) exposed to a turbulent airflow for \SI{1}{s}. Numerical results are displayed in Fig.~\ref{fig:arrest_relative_role}, which shows the evolution of the fraction of sticking/arrested/detached particles on the surface as a function of the friction velocity. It can be seen that the number of particles that halt in a region of high adhesion force increases with small values of the friction velocity before decreasing at higher velocities: this is due to the increasing number of rolling/sliding particles on the surface at small velocities which favors the migration of particles until they are either trapped in high adhesion regions or detached. As the velocity is further increased, particles trapped in high adhesion regions can be removed (by direct lift-off or migration$+$detachment motion). Nevertheless, some of the trapped particles actually remain on the surface for long times even at relatively high friction velocities (here \SI{1}{m.s^{-1}}).

In the author's opinion, deeper understanding of particle arrest is key to better understand and model long-term resuspension. Particles arrested on a surface are indeed trapped in regions where they experience high adhesion/cohesion forces and/or low hydrodynamic forces (due to sheltering). As a result, such particles will be much harder to remove from the surface and are likely to remain on the surface for a long time. Capturing their resuspension requires a precise knowledge of the force/torque distributions acting on such particles and of the occurrence of rare energetic events that can detach them (near-wall coherent structures in turbulent flows). While precise fluid simulations (such as DNS) can provide information on the near-wall structures affecting the motion of such particles, our current knowledge of the adhesion force distribution associated to arrested particles is very limited and it seems that usual distributions (Gaussian, log-normal) are not adapted to such particles.

  \section{Particle re-entrainment: dynamics after detachment from the surface}
   \label{sec:reentrain}

  \subsection{Dynamics in the near-wall turbulent region}
   \label{sec:reentrain_dynamics}
  
Once a particle is detached from a surface, its rotational and translational motion is given by:
\begin{eqnarray}
\frac{d\mb{x}_{\rm p}}{dt} & = & \mb{U}_p, \\
m_{\rm p}\frac{d\mb{U}_{\rm p}}{dt} & = & \mb{F}_{\rm f \to p} + \mb{F}_{\rm s \to p} + \mb{F}_{\rm p \to p} + \mb{F}_{\rm ext}, \\
I_{\rm p}\frac{d\mb{\Omega}_{\rm p}}{dt} & = & \mb{M}_{\rm f \to p} + \mb{M}_{\rm s \to p} + \mb{M}_{\rm p \to p} + \mb{M}_{\rm ext} 
\label{eq:part_transport_wall}
\end{eqnarray}
where the forces/torques acting on the particle are the same as those described previously in section~\ref{sec:force} with some modifications to account for the fact that particles are now at a certain distance from the surface. In particular, particle-surface interactions are expressed as a function of the inter-surface distance including both sort-ranged and long-ranged forces. As for cohesion forces, short-range interactions are often modeled using a Lennard-Jones type of potential (with both van der Waals attractive forces and Pauli repulsion forces) and adding extra short-ranged contributions such as electrostatic double-layer interactions for charged particles in liquids or steric forces for polymeric fluids \cite{henry2012towards, israelachvili2011intermolecular, liang2007interaction}.

\begin{figure}
 \centering
 \includegraphics[width=0.5\textwidth]{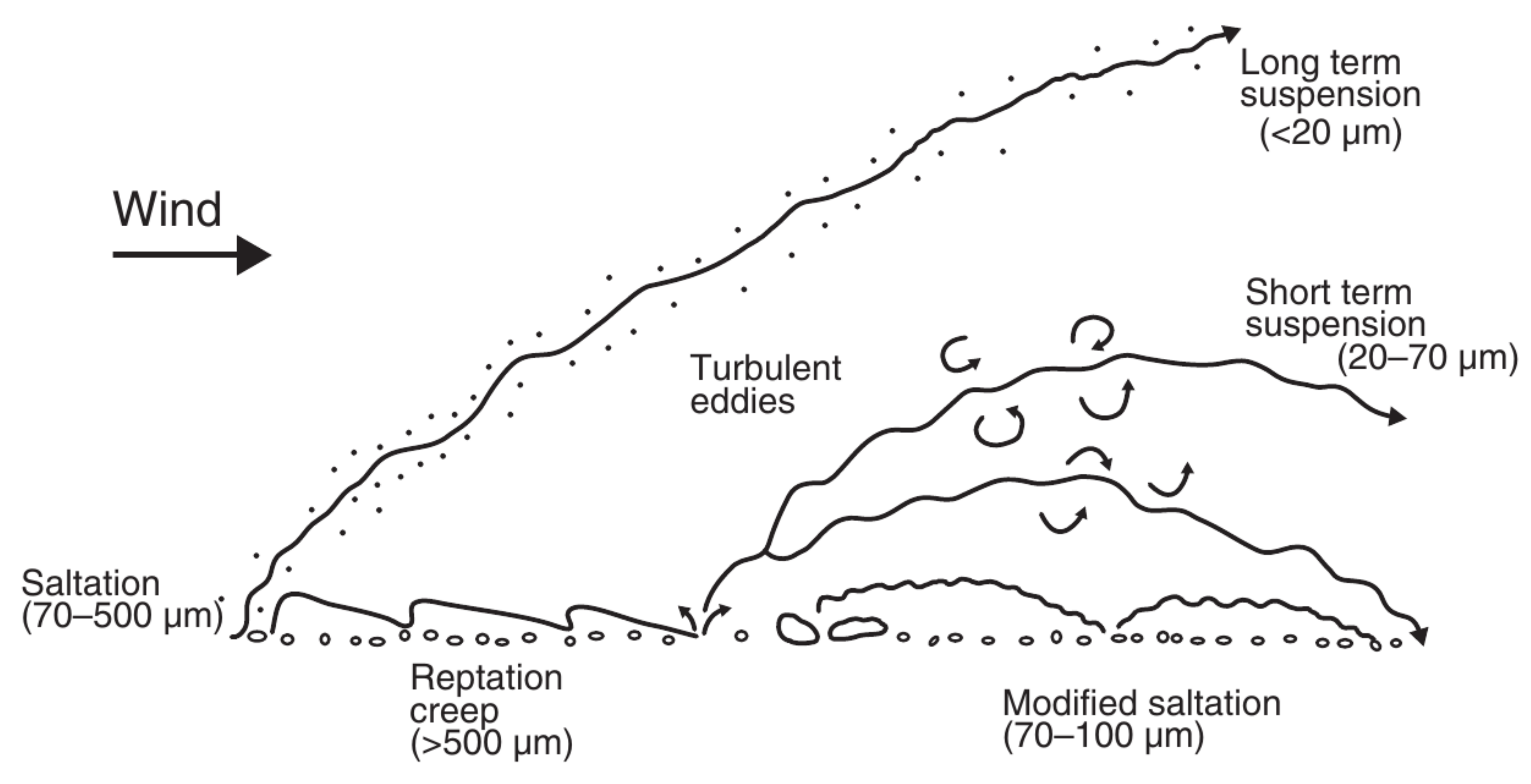}
 \caption{Schematic of the different modes of aeolian transport with the corresponding particle sizes. Reprinted from \cite{nickling2009aeolian}. Copyright 2009 Springer.}
 \label{fig:nickling_aeolian}
\end{figure}

This topic has been extensively studied in the case of sands and dust due to its impact on the environment \cite{kok2012physics}. More precisely, different modes for the transport of soil particles have been identified for aeolian transport (see also Fig.~\ref{fig:nickling_aeolian}): long-term suspension, short-term suspension, saltation, reptation and creep \cite{kok2012physics}. Saltation occurs for materials which display leap-type motion, i.e. they are detached from the surface then carried by the flow before being transported back to the surface (usually with a diameter $\sim70-500\mu m$). Upon impacting the surface, these saltating particles can eject materials present on the surface. The materials ejected by such impact can be much smaller and, due to their small size, these ejected particles will spend some time in the flow before impacting the surface again. A distinction is made between long-term suspension, i.e. particles which can remain in the atmosphere for weeks and travel thousands of kilometers (diameter $\lesssim20\mu m$ such as small sand) and short-term suspension (diameter $\sim20 - 70\mu m$). When materials ejected by saltating particles are large (diameter $\gtrsim 500\mu m$), these particles quickly settle back on the surface: motion through such very short hop-off is referred to as reptation. It should be noted that reptation thus differs from creep (where particles only roll on the surface without detaching from it). These different modes for the transport of soil particles are actually related to the time such particles spend in the atmosphere before impacting the surface again. Drawing back on the definitions introduced in section~\ref{sec:intro_term}, saltation (including short-term and long-term suspension) and reptation corresponds to specific situations of particle re-entrainment where sedimentation brings materials back to the surface within a certain finite time. 

Particle re-entrainment has also been studied extensively in the multiphase flow community (see a recent review \cite{soldati2009physics} and references therein). These studies have focused on the role of turbulent structures on the dynamics of particle transfer in the proximity of the wall. As mentioned before, boundary layers are characterized by the presence of fluid coherent structures, which can be loosely defined as regions of the fluid exhibiting a certain order in space and time. These near-wall fluid structures have been shown to significantly impact the dynamics of particles with a small relaxation time $\tau_{\rm p}$, i.e. that adapt relatively quickly to changes in the fluid velocity \cite{marchioli2002mechanisms, marchioli2003direct, soldati2009physics}. In the context of aeolian particles, such materials correspond to particles undergoing long-term suspension (i.e. sand/dust with a diameter $\lesssim20\mu m$ in an typical airflow). Their motion is highly correlated to the structures, i.e. that particles moving towards the surface tend to be correlated with `sweep' structures (i.e. with a velocity directed towards the wall) while particles moving away from the wall tend to be correlated with `ejection' structures (i.e. with a velocity directed towards the bulk of the flow). These structures thus affect the characteristics of particle impaction/deposition on surfaces: inertial particles in a sweep structure can indeed acquire enough momentum to go through the viscous sublayer and impact the surface with a relatively high kinetic energy while low-inertia particles are transported by sweep structures towards the viscous sublayer where they undergo diffusive motion (i.e. they reach the surface with a lower kinetic energy). Moreover, as mentioned previously in section~\ref{sec:incipient_initial}, ejection structures can be responsible for lifting large particles protruding from the viscous sublayer due to the high intensity associated with such events (leading to burst-type resuspension) \cite{henry2014progress, vanhout2013spatially}. More recent studies have also explored the re-entrainment of detached particles under complex conditions, such as a vertical impinging jet representative of the flow produced by a helicopter rotor \cite{wu2017particle}.

 \subsection{Direction of future studies}
 \label{sec:reentrain_future}

  \subsubsection{Multilayered systems: dynamics around complex deposits}
   \label{sec:reentrain_multilayer}

\begin{figure}
 \centering
 \includegraphics[width = 0.40\textwidth]{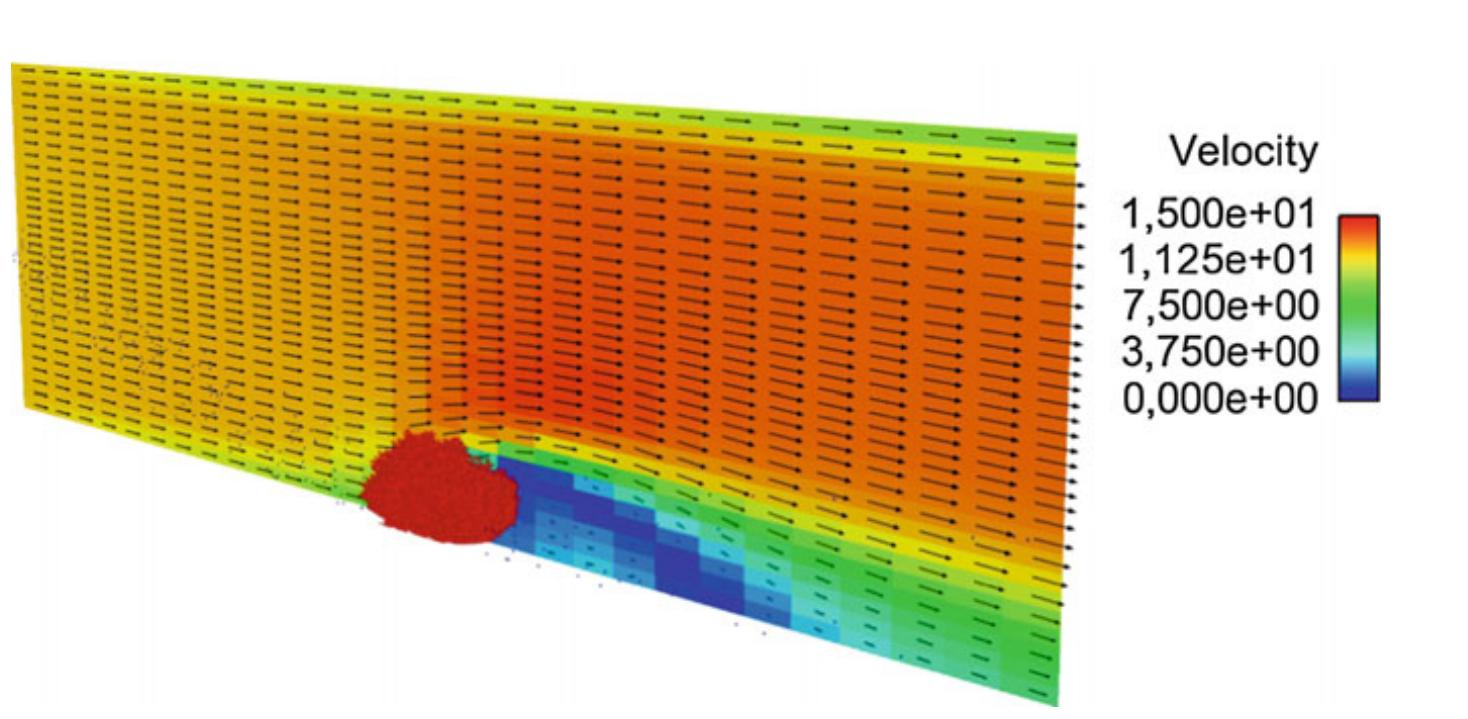}
 \caption{Fluid flow around a deposit. Results obtained from a numerical simulation with two-way coupling. Reprinted from \cite{caruyer2016stochastic}. Copyright 2016 Springer.}
 \label{fig:flow_deposit}
\end{figure}

Even though the motion of particles after detachment is fairly well characterized in the case of monolayered systems, our understanding of the dynamics of particles resuspended from multilayered systems still remains limited. Most experimental and numerical studies of particles in wall-bounded turbulent flows have been focused on the case of a smooth surface (from a hydrodynamic point of view at least). Yet, as a deposit grows on a surface, its size can becomes comparable to typical hydrodynamic scales and thus has an impact on the flow around it: this is illustrated in Fig.~\ref{fig:flow_deposit} where the recirculation can be seen in the wake of the deposit. Recent studies have shown that roughness effects on turbulent boundary layers can be characterized by two parameters (more details can be found in \cite{jimenez2004turbulent}: the roughness Reynolds number, which measures the extent of roughness interference in the viscous buffer layer, and the blockage ratio (ratio of the boundary layer thickness to the roughness height), which quantifies the effect of roughness on the logarithmic layer. It was shown that the buffer layer is modified even for roughness elements as small as a few all units. Meanwhile, the logarithmic layer is impacted by roughness when the blockage ratio is higher than $40-80$ (\SI{\sim100}{\mu m} for water flowing at \SI{1}{m.s^{-1}}) \cite{jimenez2004turbulent}. When deposits form large-scale structures on the surface (i.e. high values of the blockage ratio), roughness not only affects small-scale local characteristics of the fluid but also its large-scale structures: in that case, the flow is actually best described by flows over obstacles. The flow over such rough surfaces is also a complex function of the roughness size and surface density (i.e. the spacing between two consecutive features): for instance, in aeolian transport, the formation and dynamics of self-organized structures such as wind ripples (dunes) results from the complex interaction between sediments and wind \cite{kok2012physics, lancaster1994dune}.

Despite these recent advances, the effect of rough surfaces on turbulent boundary layers is far from being understood. Among all the open issues that require further investigation (see \cite{jimenez2004turbulent}), the effect of roughness on the fluid structures is crucial in the context of particle resuspension. In particular, it remains to be seen how the coherent structures present in boundary layers are affected by surface roughness due to their key role in the dynamics of medium-sized particles near the wall. Yet, the main difficulty is related to the wide range of scales displayed by such deposits: accumulation of particles can indeed results in large-scale features on the surface while local (small-scale) variations of the deposit height are proportional to the particle size. Therefore, their effect on the spatial and temporal characteristics of coherent structures is not straightforward and can be very system-dependent. In the author's opinion, insights into the local flow structure around complex deposits can be investigated using time-resolved particle image velocimetry (PIV) and particle tracking velocimetry \cite{vanhout2013spatially,traugott2017experimental}. This technique has been indeed recently shown to provide precise measurements of the resuspension and saltation of suspended particles (\SI{583}{\mu m} in diameter) in a turbulent water channel flow \cite{vanhout2013spatially}. The interest of this technique is that it can not only provide information on the effect of deposits on near-wall coherent structures but also on the role of such structures in particle resuspension.

  \subsubsection{Saltation: outcome of re-impact on the surface}
   \label{sec:reentrain_redeposit}

\begin{figure}
 \centering
 \includegraphics[width=0.45\textwidth]{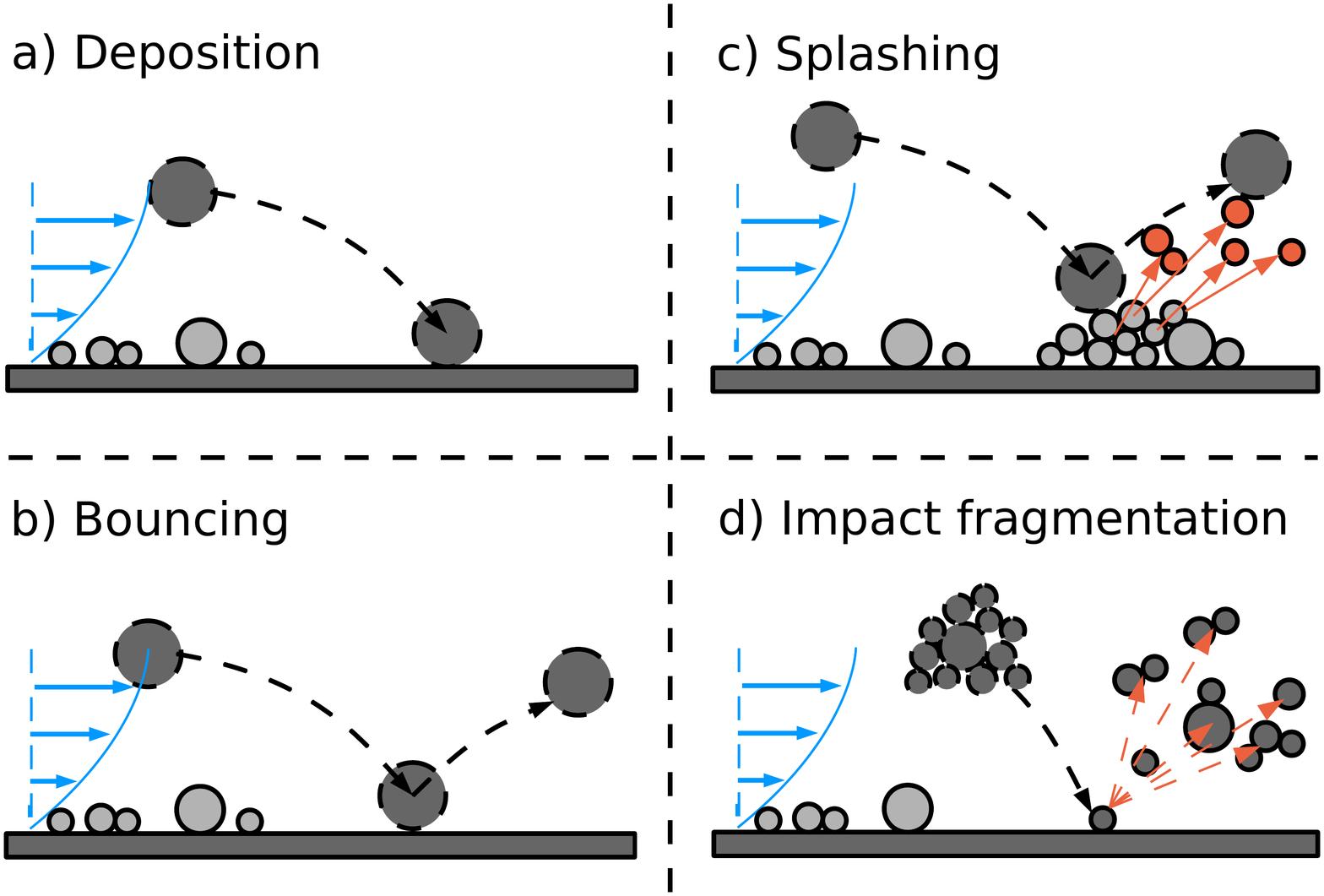}
 \caption{Sketch showing the various possible outcomes after a saltating particle impacts a surface.}
 \label{fig:sketch_saltation}
\end{figure}

Another limitation in our current understanding of particle re-entrainment is related to saltation: saltating particles impact the surface after being re-entrained in the fluid for a certain time (which is inversely proportional to their mass). Upon colliding with the surface, various outcomes are possible (see also Fig.~\ref{fig:sketch_saltation}):
\begin{itemize}
 \item re-deposition takes place when the saltating particle adhere again on the surface. This usually occurs when the impact energy is dissipated during the collision process. In the case of solid particles, dissipation is often related to elasto-plastic deformations of surface (interested readers are referred to the book \cite{stronge2004impact} for details on impact mechanics). This process is rendered more complex when a saltating particle impacts an existing multilayered deposit since the force/energy due to the impact will propagate within the complex network of particles and dissipate either through local deformations or deposit restructuration (see also \cite{bourrier2008physical}).
 \item rebound occurs when the impact energy is not fully dissipated by non-conservative forces (such as friction, viscous or plastic processes) and part of this energy is actually restituted to the particle (more details in \cite{stronge2004impact}). This also explains the distinction between elastic collisions, i.e. where no energy is dissipated during the collision, and inelastic collisions, i.e. where dissipation occurs. As for re-deposition, bouncing on multilayered deposits does not only depends on the first energy exchange between the saltating particle and the deposit but also on the shock-wave propagation through it and reflection at the surface \cite{bourrier2008physical}.
 \item splashing occurs when particles impact the surface with a kinetic energy high enough to move some of the materials present on the surface. This splashing has been extensively studied for sand/dust particles in the context of aeolian transport since it was shown to increase the detachment rate of large aerosols (diameter $\sim$ \SI{100}{\mu m}) by a factor two \cite{fairchild1982wind}. These studies have also characterized what is the effect of the impact angle and the impact velocity on splashing, and especially how it affects the number of splashed elements and their angle of ejection (more details can be found in \cite{henry2014progress}). Yet, one of the difficulties that occurs in the development of accurate theories/models for splashing is related to the possible chain reaction that results from splashing: a particle impacting a surface can lead to the resuspension of several other particles, each of those potentially playing the role of a new splashing particles. Such chain reactions are in some sense similar to the formation of loose-snow avalanche which start at a specific region and spread out as snow moves down the slope \cite{mcclung2006avalanche}.
 \item Impact fragmentation corresponds to the case of aggregates (composed of several particles held together by cohesion forces) which fragment upon impact on the surface. Each fragment can either deposit on the surface or bounce depending on its kinetic energy and the interaction with the surface. Impact fragmentation has been observed for structures made of colloidal particles and these studies have shown that it depends on the impact velocity, impact angle, aggregate structure and cohesion forces (see in particular \cite{ihalainen2014break} and references therein).
\end{itemize}

The outcome of the collision of a saltating particle on a surface thus depends on a number of parameters including those related to the impact (velocity, angle), to the particle (deformability) and to the surface (deformability, deposit structure). While the outcome of a saltating particle on a surface is well understood, the impact on multilayered structures still requires further characterizations due to the role of the complex deposit morphology. To go beyond the current understanding of saltation, there is thus a need for new detailed and systematic experimental and numerical studies. In particular, the evolution of the deposit morphology as a function of fluid (velocity, etc.) and surface properties needs to be further characterized for small colloidal particles. In the author's opinion, this could be achieved using microfluidic experiments with a suspension of small colloids as in \cite{mustin2010deposition}. Then, the effect of the deposit structure on splashing needs to be asserted experimentally. Such systematic experiments would also allow to assess the relative importance of the various outcomes (i.e. re-deposition, bouncing, splashing and impact fragmentation) as a function of the size of the saltating particle. It is indeed not evident whether impact-fragmentation plays a role in the case of large suspended particles while splashing effects can be negligible for colloids (due to their small kinetic energy). For example, recent studies in spray drying systems have shown that the impact of aggregates on surfaces leads to the break-up of clusters on the wall \cite{francia2017agglomeration}. Besides, the effect of deposit restructuration needs to be further assessed in such experiments especially for colloidal particles. Re-arrangement has been indeed confirmed in the case of suspended particles: for instance, it has been estimated that density increase in alpine snow due to grain rearrangements occurs mostly through the slow creep motion of snow caused by the weight of layers above while mechanical effects only contribute to a very small portion of density increase (\cite{mcclung2006avalanche}. Yet, it remains to be seen whether such re-arrangements also occur in the case of smaller particles (where contact force are more predominant).

  \subsubsection{Particles shape: non-sphericity and flexibility}
   \label{sec:reentrain_nonspherical}

As mentioned previously in section~\ref{sec:incipient_future_nonsphericity}, real particles often have a non-spherical shape and some particles even deform (see section~\ref{sec:incipient_future_flexibility}). As a result, both the dynamics of the center of mass of the particle and the orientation of the particle need to be studied. Recent studies have confirmed that small particles behave like tracers with translational motion matching the fluid flow and rotational motion matching the vorticity field. With the recent advances in experimental techniques and DNS simulations, rapid progress has also been made in the understanding of their dynamics and orientation in the near-wall region \cite{voth2017anisotropic}. In particular, several studies have investigated the dynamics of rigid spheroidal particles with various aspect ratio near the wall (more details in \cite{voth2017anisotropic} and references therein): the alignment of prolate ellipsoids was shown to follow the streamwise direction while the alignment of oblate ellipsoids depends on their inertia (with tracers aligning perpendicularly to the wall and heavy disks aligning in the spanwise direction). Yet, even with the recent advances in experimental and numerical techniques, the dynamics of flexible particles (such as fibers or polymers) still remains partially understood. This is due to the large number of parameters needed to describe such systems (due to partial folding of such elongated particles). This additional complexity calls for further experimental studies and also for numerical simulations capable of handling arbitrarily shaped particles \cite{voth2017anisotropic}.

Besides the limitations related to the dynamics of non-spherical particles in the near-wall region, there is an ongoing need to characterize the effect of non-sphericity on the outcome of a collision with a surface. The adhesion of non-spherical particles has indeed been characterized mostly for rigid spheroids \cite{brambilla2017adhesion} but not for arbitrarily shaped particles. In particular, a flexible particle exhibits more complex interactions with a surface since it can be partially adhering to a surface (see also Fig.~\ref{fig:adhesion_fiber_grebikova}). As a result, one expects that its impact on a surface can produce a range of possible outcome that need to be characterized both experimentally and theoretically.
  
\section{Conclusion}
 \label{sec:conclusion}

This review provides a general framework and terminology to address particle resuspension from surfaces, which is described in terms of three underlying mechanisms: the incipient motion of particles (i.e. how sticking particles are set in motion), the migration on the surface (through rolling or sliding motion) until detachment, the re-entrainment in the flow (i.e. their motion after detachment). For each of the three mechanisms, existing results were briefly summarized to highlight the underlying physical notions before listing the limitations in our current understanding while providing possible suggestions on how to address these issues in the near future using recent techniques.

First, the incipient motion of particles was shown to depend on the balance between forces/torques acting on particles. This balance can lead to three types of motion: direct lift-off, rolling or sliding motion. While the overall effect of the various forces at play appears to be well understood, there is an ongoing need to get more details on the exact distribution of these forces in complex environments. This includes in particular the case of adhesion forces between rough surfaces and cohesion forces in multilayered deposits since the highest values of these forces (tail of the distribution) directly impact long-term resuspension. Second, the migration of particles on a surface occurs through rolling or sliding motion until the particle is detached from the surface (either through direct lift-off or through rocking events). The role of particle migration in the overall resuspension phenomena is still not fully understood: there is indeed a need to characterize the spatial and temporal scales associated with migration before a particle actually detach from the surface. Besides, particle migration can lead to the arrest of particle in high adhesion regions. Such particles can remain trapped in these regions for very long times. It is thus crucial to investigate this issue in the near future since it can have a profound impact on long-term resuspension. Third, particle re-entrainment has been shown to display different features (short-term or long-term suspension, saltation, reptation) which depend on both the particles and near-wall fluid structures. Yet, further studies are needed to assess the relative importance between the various outcomes of saltating particles impacting a surface (i.e. re-deposition, bouncing, splashing, impact fragmentation) and how it changes with particles and fluid properties. Last, future studies are also needed to investigate the role of particle shape on resuspension: more complex dynamics are indeed expected for rigid non-spherical particles or flexible particles due to the effect of the particle shape on both adhesion and hydrodynamic forces.

In the author's opinion, such refined characterizations of the features associated to particle resuspension will help designing better modeling approaches. For instance, there is currently no general model that actually takes into account all the aspects of particle resuspension (i.e. incipient motion, migration and re-entrainment including splashing and impact fragmentation effects). Besides, progress in our understanding of particle resuspension will also provide significant advances in the field of particle fouling, since the growth of materials on a surface is controlled by the balance between deposition and resuspension \cite{francia2017agglomeration, matsusaka2015high, minier2016particles}.

\section*{Acknowledgements}
The author would like to express special thanks to Jean-Pierre Minier for fruitful discussions and constructive remarks that helped improve this manuscript. The author also acknowledges the support of the EU COST Action MP1305 ``Flowing Matter''.

\bibliography{Bibliography}

\begin{thebibliography}{91}
\expandafter\ifx\csname natexlab\endcsname\relax\def\natexlab#1{#1}\fi
\providecommand{\url}[1]{\texttt{#1}}
\providecommand{\href}[2]{#2}
\providecommand{\path}[1]{#1}
\providecommand{\DOIprefix}{doi:}
\providecommand{\ArXivprefix}{arXiv:}
\providecommand{\URLprefix}{URL: }
\providecommand{\Pubmedprefix}{pmid:}
\providecommand{\doi}[1]{\href{http://dx.doi.org/#1}{\path{#1}}}
\providecommand{\Pubmed}[1]{\href{pmid:#1}{\path{#1}}}
\providecommand{\bibinfo}[2]{#2}
\ifx\xfnm\relax \def\xfnm[#1]{\unskip,\space#1}\fi
\bibitem[{Abd-Elhady et~al.(2007)Abd-Elhady, Clevers, Adriaans, Rindt, Wijers
  \& Van~Steenhoven}]{abd2007influence}
\bibinfo{author}{Abd-Elhady, M.}, \bibinfo{author}{Clevers, S.},
  \bibinfo{author}{Adriaans, T.}, \bibinfo{author}{Rindt, C.},
  \bibinfo{author}{Wijers, J.}, \& \bibinfo{author}{Van~Steenhoven, A.}
  (\bibinfo{year}{2007}).
\newblock \bibinfo{title}{Influence of sintering on the growth rate of
  particulate fouling layers}.
\newblock {\it \bibinfo{journal}{International Journal of Heat and Mass
  Transfer}\/},  {\it \bibinfo{volume}{50}\/}, \bibinfo{pages}{196--207}.
  \URLprefix \url{https://doi.org/10.1016/j.ijheatmasstransfer.2006.06.018}.
\bibitem[{Agudo et~al.(2017)Agudo, Luzi, Han, Hwang, Lee \&
  Wierschem}]{agudo2017detection}
\bibinfo{author}{Agudo, J.}, \bibinfo{author}{Luzi, G.}, \bibinfo{author}{Han,
  J.}, \bibinfo{author}{Hwang, M.}, \bibinfo{author}{Lee, J.}, \&
  \bibinfo{author}{Wierschem, A.} (\bibinfo{year}{2017}).
\newblock \bibinfo{title}{Detection of particle motion using image processing
  with particular emphasis on rolling motion}.
\newblock {\it \bibinfo{journal}{Review of Scientific Instruments}\/},  {\it
  \bibinfo{volume}{88}\/}, \bibinfo{pages}{051805}. \URLprefix
  \url{https://doi.org/10.1063/1.4983054}.
\bibitem[{Audry et~al.(2009)Audry, Ramos \& Charlaix}]{audry2009adhesion}
\bibinfo{author}{Audry, M.-C.}, \bibinfo{author}{Ramos, S.}, \&
  \bibinfo{author}{Charlaix, E.} (\bibinfo{year}{2009}).
\newblock \bibinfo{title}{Adhesion between highly rough alumina surfaces: an
  atomic force microscope study}.
\newblock {\it \bibinfo{journal}{Journal of colloid and interface science}\/},
  {\it \bibinfo{volume}{331}\/}, \bibinfo{pages}{371--378}. \URLprefix
  \url{https://doi.org/10.1016/j.jcis.2008.11.050}.
\bibitem[{Bagnold(2012)}]{bagnold2012physics}
\bibinfo{author}{Bagnold, R.~A.} (\bibinfo{year}{2012}).
\newblock {\it \bibinfo{title}{The physics of blown sand and desert dunes}\/}.
\newblock \bibinfo{publisher}{Courier Corporation}.
\bibitem[{Banerjee et~al.(2011)Banerjee, Pangule \&
  Kane}]{banerjee2011antifouling}
\bibinfo{author}{Banerjee, I.}, \bibinfo{author}{Pangule, R.~C.}, \&
  \bibinfo{author}{Kane, R.~S.} (\bibinfo{year}{2011}).
\newblock \bibinfo{title}{Antifouling coatings: recent developments in the
  design of surfaces that prevent fouling by proteins, bacteria, and marine
  organisms}.
\newblock {\it \bibinfo{journal}{Advanced Materials}\/},  {\it
  \bibinfo{volume}{23}\/}, \bibinfo{pages}{690--718}. \URLprefix
  \url{https://doi.org/10.1080/08927010802256117}.
\bibitem[{Basset(1888)}]{basset1888treatise}
\bibinfo{author}{Basset, A.~B.} (\bibinfo{year}{1888}).
\newblock {\it \bibinfo{title}{A treatise on hydrodynamics: with numerous
  examples}\/} volume~\bibinfo{volume}{2}.
\newblock \bibinfo{publisher}{Deighton, Bell and Company}.
\bibitem[{Beach et~al.(2002)Beach, Tormoen, Drelich \& Han}]{beach2002pull}
\bibinfo{author}{Beach, E.}, \bibinfo{author}{Tormoen, G.},
  \bibinfo{author}{Drelich, J.}, \& \bibinfo{author}{Han, R.}
  (\bibinfo{year}{2002}).
\newblock \bibinfo{title}{Pull-off force measurements between rough surfaces by
  atomic force microscopy}.
\newblock {\it \bibinfo{journal}{Journal of Colloid and Interface Science}\/},
  {\it \bibinfo{volume}{247}\/}, \bibinfo{pages}{84--99}. \URLprefix
  \url{https://doi.org/10.1006/jcis.2001.8126}.
\bibitem[{Boor et~al.(2013)Boor, Siegel \& Novoselac}]{boor2013monolayer}
\bibinfo{author}{Boor, B.~E.}, \bibinfo{author}{Siegel, J.~A.}, \&
  \bibinfo{author}{Novoselac, A.} (\bibinfo{year}{2013}).
\newblock \bibinfo{title}{Monolayer and multilayer particle deposits on hard
  surfaces: literature review and implications for particle resuspension in the
  indoor environment}.
\newblock {\it \bibinfo{journal}{Aerosol Science and Technology}\/},  {\it
  \bibinfo{volume}{47}\/}, \bibinfo{pages}{831--847}. \URLprefix
  \url{https://doi.org/10.1080/02786826.2013.794928}.
\bibitem[{Bourrier et~al.(2008)Bourrier, Nicot \& Darve}]{bourrier2008physical}
\bibinfo{author}{Bourrier, F.}, \bibinfo{author}{Nicot, F.}, \&
  \bibinfo{author}{Darve, F.} (\bibinfo{year}{2008}).
\newblock \bibinfo{title}{Physical processes within a 2d granular layer during
  an impact}.
\newblock {\it \bibinfo{journal}{Granular Matter}\/},  {\it
  \bibinfo{volume}{10}\/}, \bibinfo{pages}{415--437}. \URLprefix
  \url{https://doi.org/10.1007/s10035-008-0108-0}.
\bibitem[{Braaten(1994)}]{braaten1994wind}
\bibinfo{author}{Braaten, D.~A.} (\bibinfo{year}{1994}).
\newblock \bibinfo{title}{Wind tunnel experiments of large particle
  reentrainment-deposition and development of large particle scaling
  parameters}.
\newblock {\it \bibinfo{journal}{Aerosol Science and Technology}\/},  {\it
  \bibinfo{volume}{21}\/}, \bibinfo{pages}{157--169}. \URLprefix
  \url{https://doi.org/10.1080/02786829408959705}.
\bibitem[{Brambilla et~al.(2017)Brambilla, Speckart \&
  Brown}]{brambilla2017adhesion}
\bibinfo{author}{Brambilla, S.}, \bibinfo{author}{Speckart, S.}, \&
  \bibinfo{author}{Brown, M.~J.} (\bibinfo{year}{2017}).
\newblock \bibinfo{title}{Adhesion and aerodynamic forces for the resuspension
  of non-spherical particles in outdoor environments}.
\newblock {\it \bibinfo{journal}{Journal of Aerosol Science}\/},  {\it
  \bibinfo{volume}{112}\/}, \bibinfo{pages}{52--67}. \URLprefix
  \url{https://doi.org/10.1016/j.jaerosci.2017.07.006}.
\bibitem[{Caruyer et~al.(2016)Caruyer, Minier, Guingo \&
  Henry}]{caruyer2016stochastic}
\bibinfo{author}{Caruyer, C.}, \bibinfo{author}{Minier, J.-P.},
  \bibinfo{author}{Guingo, M.}, \& \bibinfo{author}{Henry, C.}
  (\bibinfo{year}{2016}).
\newblock \bibinfo{title}{A stochastic model for particle deposition in
  turbulent flows and clogging effects}.
\newblock In {\it \bibinfo{booktitle}{Advances in Hydroinformatics}\/} (pp.
  \bibinfo{pages}{597--612}).
\newblock \bibinfo{publisher}{Springer}.
\newblock \URLprefix \url{https://doi.org/10.1007/978-981-287-615-7$\_$40}.
\bibitem[{Clift et~al.(2005)Clift, Grace \& Weber}]{clift2005bubbles}
\bibinfo{author}{Clift, R.}, \bibinfo{author}{Grace, J.~R.}, \&
  \bibinfo{author}{Weber, M.~E.} (\bibinfo{year}{2005}).
\newblock {\it \bibinfo{title}{Bubbles, drops, and particles}\/}.
\newblock \bibinfo{publisher}{Courier Corporation}.
\bibitem[{Coleman \& Nikora(2008)}]{coleman2008unifying}
\bibinfo{author}{Coleman, S.}, \& \bibinfo{author}{Nikora, V.}
  (\bibinfo{year}{2008}).
\newblock \bibinfo{title}{A unifying framework for particle entrainment}.
\newblock {\it \bibinfo{journal}{Water resources research}\/},  {\it
  \bibinfo{volume}{44}\/}. \URLprefix
  \url{http://dx.doi.org/10.1029/2007WR006363}.
\bibitem[{De~Vet et~al.(2014)De~Vet, Merrison, Mittelmeijer-Hazeleger, Van~Loon
  \& Cammeraat}]{de2014effects}
\bibinfo{author}{De~Vet, S.}, \bibinfo{author}{Merrison, J.},
  \bibinfo{author}{Mittelmeijer-Hazeleger, M.}, \bibinfo{author}{Van~Loon, E.},
  \& \bibinfo{author}{Cammeraat, L.} (\bibinfo{year}{2014}).
\newblock \bibinfo{title}{Effects of rolling on wind-induced detachment
  thresholds of volcanic glass on mars}.
\newblock {\it \bibinfo{journal}{Planetary and Space Science}\/},  {\it
  \bibinfo{volume}{103}\/}, \bibinfo{pages}{205--218}. \URLprefix
  \url{https://doi.org/10.1016/j.pss.2014.07.012}.
\bibitem[{Derjaguin et~al.(1975)Derjaguin, Muller \&
  Toporov}]{derjaguin1975effect}
\bibinfo{author}{Derjaguin, B.~V.}, \bibinfo{author}{Muller, V.~M.}, \&
  \bibinfo{author}{Toporov, Y.~P.} (\bibinfo{year}{1975}).
\newblock \bibinfo{title}{Effect of contact deformations on the adhesion of
  particles}.
\newblock {\it \bibinfo{journal}{Journal of Colloid and interface science}\/},
  {\it \bibinfo{volume}{53}\/}, \bibinfo{pages}{314--326}. \URLprefix
  \url{https://doi.org/10.1016/0021-9797(75)90018-1}.
\bibitem[{Duru \& Hallez(2015)}]{duru2015three}
\bibinfo{author}{Duru, P.}, \& \bibinfo{author}{Hallez, Y.}
  (\bibinfo{year}{2015}).
\newblock \bibinfo{title}{A three-step scenario involved in particle capture on
  a pore edge}.
\newblock {\it \bibinfo{journal}{Langmuir}\/},  {\it \bibinfo{volume}{31}\/},
  \bibinfo{pages}{8310--8317}. \URLprefix
  \url{http://pubs.acs.org/doi/abs/10.1021/acs.langmuir.5b01298}.
\bibitem[{Eichenlaub et~al.(2004)Eichenlaub, Gelb \&
  Beaudoin}]{eichenlaub2004roughness}
\bibinfo{author}{Eichenlaub, S.}, \bibinfo{author}{Gelb, A.}, \&
  \bibinfo{author}{Beaudoin, S.} (\bibinfo{year}{2004}).
\newblock \bibinfo{title}{Roughness models for particle adhesion}.
\newblock {\it \bibinfo{journal}{Journal of colloid and interface science}\/},
  {\it \bibinfo{volume}{280}\/}, \bibinfo{pages}{289--298}. \URLprefix
  \url{https://doi.org/10.1016/j.jcis.2004.08.017}.
\bibitem[{Elghobashi(1994)}]{elghobashi1994predicting}
\bibinfo{author}{Elghobashi, S.} (\bibinfo{year}{1994}).
\newblock \bibinfo{title}{On predicting particle-laden turbulent flows}.
\newblock {\it \bibinfo{journal}{Applied scientific research}\/},  {\it
  \bibinfo{volume}{52}\/}, \bibinfo{pages}{309--329}. \URLprefix
  \url{https://doi.org/10.1007/BF00936835}.
\bibitem[{Fairchild \& Tillery(1982)}]{fairchild1982wind}
\bibinfo{author}{Fairchild, C.~I.}, \& \bibinfo{author}{Tillery, M.~I.}
  (\bibinfo{year}{1982}).
\newblock \bibinfo{title}{Wind tunnel measurements of the resuspension of ideal
  particles}.
\newblock {\it \bibinfo{journal}{Atmospheric Environment (1967)}\/},  {\it
  \bibinfo{volume}{16}\/}, \bibinfo{pages}{229--238}. \URLprefix
  \url{https://doi.org/10.1016/0004-6981(82)90437-1}.
\bibitem[{Fan et~al.(1997)Fan, Soltani, Ahmadi \& Hart}]{fan1997flow}
\bibinfo{author}{Fan, F.-G.}, \bibinfo{author}{Soltani, M.},
  \bibinfo{author}{Ahmadi, G.}, \& \bibinfo{author}{Hart, S.~C.}
  (\bibinfo{year}{1997}).
\newblock \bibinfo{title}{Flow-induced resuspension of rigid-link fibers from
  surfaces}.
\newblock {\it \bibinfo{journal}{Aerosol science and technology}\/},  {\it
  \bibinfo{volume}{27}\/}, \bibinfo{pages}{97--115}. \URLprefix
  \url{https://doi.org/10.1080/02786829708965460}.
\bibitem[{Francia et~al.(2015)Francia, Mart{\'\i}n, Bayly \&
  Simmons}]{francia2015role}
\bibinfo{author}{Francia, V.}, \bibinfo{author}{Mart{\'\i}n, L.},
  \bibinfo{author}{Bayly, A.~E.}, \& \bibinfo{author}{Simmons, M.~J.}
  (\bibinfo{year}{2015}).
\newblock \bibinfo{title}{The role of wall deposition and re-entrainment in
  swirl spray dryers}.
\newblock {\it \bibinfo{journal}{AIChE Journal}\/},  {\it
  \bibinfo{volume}{61}\/}, \bibinfo{pages}{1804--1821}. \URLprefix
  \url{http://dx.doi.org/10.1002/aic.14767}.
\bibitem[{Francia et~al.(2017)Francia, Mart{\'\i}n, Bayly \&
  Simmons}]{francia2017agglomeration}
\bibinfo{author}{Francia, V.}, \bibinfo{author}{Mart{\'\i}n, L.},
  \bibinfo{author}{Bayly, A.~E.}, \& \bibinfo{author}{Simmons, M.~J.}
  (\bibinfo{year}{2017}).
\newblock \bibinfo{title}{Agglomeration during spray drying: Airborne clusters
  or breakage at the walls?}
\newblock {\it \bibinfo{journal}{Chemical Engineering Science}\/},  {\it
  \bibinfo{volume}{162}\/}, \bibinfo{pages}{284--299}. \URLprefix
  \url{https://doi.org/10.1016/j.ces.2016.12.033}.
\bibitem[{Friess \& Yadigaroglu(2002)}]{friess2002modelling}
\bibinfo{author}{Friess, H.}, \& \bibinfo{author}{Yadigaroglu, G.}
  (\bibinfo{year}{2002}).
\newblock \bibinfo{title}{Modelling of the resuspension of particle clusters
  from multilayer aerosol deposits with variable porosity}.
\newblock {\it \bibinfo{journal}{Journal of Aerosol Science}\/},  {\it
  \bibinfo{volume}{33}\/}, \bibinfo{pages}{883--906}. \URLprefix
  \url{https://doi.org/10.1016/S0021-8502(02)00049-6}.
\bibitem[{Gatignol(1983)}]{gatignol1983faxen}
\bibinfo{author}{Gatignol, R.} (\bibinfo{year}{1983}).
\newblock \bibinfo{title}{The fax{\'e}n formulas for a rigid particle in an
  unsteady non-uniform stokes-flow}.
\newblock {\it \bibinfo{journal}{Journal de M{\'e}canique th{\'e}orique et
  appliqu{\'e}e}\/},  {\it \bibinfo{volume}{2}\/}, \bibinfo{pages}{143--160}.
\bibitem[{G{\"o}tzinger \& Peukert(2004)}]{gotzinger2004particle}
\bibinfo{author}{G{\"o}tzinger, M.}, \& \bibinfo{author}{Peukert, W.}
  (\bibinfo{year}{2004}).
\newblock \bibinfo{title}{Particle adhesion force distributions on rough
  surfaces}.
\newblock {\it \bibinfo{journal}{Langmuir}\/},  {\it \bibinfo{volume}{20}\/},
  \bibinfo{pages}{5298--5303}. \URLprefix
  \url{http://pubs.acs.org/doi/abs/10.1021/la049914f}.
\bibitem[{Grado{\'n}(2009)}]{gradon2009resuspension}
\bibinfo{author}{Grado{\'n}, L.} (\bibinfo{year}{2009}).
\newblock \bibinfo{title}{Resuspension of particles from surfaces:
  technological, environmental and pharmaceutical aspects}.
\newblock {\it \bibinfo{journal}{Advanced Powder Technology}\/},  {\it
  \bibinfo{volume}{20}\/}, \bibinfo{pages}{17--28}. \URLprefix
  \url{https://doi.org/10.1016/j.apt.2008.10.009}.
\bibitem[{Greb{\'\i}kov{\'a} et~al.(2017)Greb{\'\i}kov{\'a}, Gojzewski,
  Kieviet, Klein~Gunnewiek \& Vancso}]{grebikova2017pulling}
\bibinfo{author}{Greb{\'\i}kov{\'a}, L.}, \bibinfo{author}{Gojzewski, H.},
  \bibinfo{author}{Kieviet, B.}, \bibinfo{author}{Klein~Gunnewiek, M.}, \&
  \bibinfo{author}{Vancso, G.} (\bibinfo{year}{2017}).
\newblock \bibinfo{title}{Pulling angle-dependent force microscopy}.
\newblock {\it \bibinfo{journal}{Review of scientific instruments}\/},  {\it
  \bibinfo{volume}{88}\/}, \bibinfo{pages}{033705}. \URLprefix
  \url{https://doi.org/10.1063/1.4978452}.
\bibitem[{Guingo \& Minier(2008)}]{guingo2008new}
\bibinfo{author}{Guingo, M.}, \& \bibinfo{author}{Minier, J.-P.}
  (\bibinfo{year}{2008}).
\newblock \bibinfo{title}{A new model for the simulation of particle
  resuspension by turbulent flows based on a stochastic description of wall
  roughness and adhesion forces}.
\newblock {\it \bibinfo{journal}{Journal of Aerosol Science}\/},  {\it
  \bibinfo{volume}{39}\/}, \bibinfo{pages}{957--973}. \URLprefix
  \url{https://doi.org/10.1016/j.jaerosci.2008.06.007}.
\bibitem[{Hall(1988)}]{hall1988measurements}
\bibinfo{author}{Hall, D.} (\bibinfo{year}{1988}).
\newblock \bibinfo{title}{Measurements of the mean force on a particle near a
  boundary in turbulent flow}.
\newblock {\it \bibinfo{journal}{Journal of Fluid Mechanics}\/},  {\it
  \bibinfo{volume}{187}\/}, \bibinfo{pages}{451--466}. \URLprefix
  \url{https://doi.org/10.1017/S0022112088000515}.
\bibitem[{Hamaker(1937)}]{hamaker1937london}
\bibinfo{author}{Hamaker, H.} (\bibinfo{year}{1937}).
\newblock \bibinfo{title}{The london—van der waals attraction between
  spherical particles}.
\newblock {\it \bibinfo{journal}{physica}\/},  {\it \bibinfo{volume}{4}\/},
  \bibinfo{pages}{1058--1072}. \URLprefix
  \url{https://doi.org/10.1016/S0031-8914(37)80203-7}.
\bibitem[{Henry \& Minier(2014{\natexlab{a}})}]{henry2014progress}
\bibinfo{author}{Henry, C.}, \& \bibinfo{author}{Minier, J.-P.}
  (\bibinfo{year}{2014}{\natexlab{a}}).
\newblock \bibinfo{title}{Progress in particle resuspension from rough surfaces
  by turbulent flows}.
\newblock {\it \bibinfo{journal}{Progress in Energy and Combustion Science}\/},
   {\it \bibinfo{volume}{45}\/}, \bibinfo{pages}{1--53}. \URLprefix
  \url{https://doi.org/10.1016/j.pecs.2014.06.001}.
\bibitem[{Henry \& Minier(2014{\natexlab{b}})}]{henry2014stochastic}
\bibinfo{author}{Henry, C.}, \& \bibinfo{author}{Minier, J.-P.}
  (\bibinfo{year}{2014}{\natexlab{b}}).
\newblock \bibinfo{title}{A stochastic approach for the simulation of particle
  resuspension from rough substrates: Model and numerical implementation}.
\newblock {\it \bibinfo{journal}{Journal of Aerosol Science}\/},  {\it
  \bibinfo{volume}{77}\/}, \bibinfo{pages}{168--192}. \URLprefix
  \url{https://doi.org/10.1016/j.jaerosci.2014.08.005}.
\bibitem[{Henry \& Minier(2018)}]{henry2018colloidal}
\bibinfo{author}{Henry, C.}, \& \bibinfo{author}{Minier, J.-P.}
  (\bibinfo{year}{2018}).
\newblock \bibinfo{title}{Colloidal particle resuspension: on the need for
  refined characterisation of surface roughness}.
\newblock {\it \bibinfo{journal}{Journal of Aerosol Science}\/},  {\it
  \bibinfo{volume}{118}\/}, \bibinfo{pages}{1--13}. \URLprefix
  \url{https://doi.org/10.1016/j.jaerosci.2018.01.005}.
\bibitem[{Henry et~al.(2012)Henry, Minier \& Lef{\`e}vre}]{henry2012towards}
\bibinfo{author}{Henry, C.}, \bibinfo{author}{Minier, J.-P.}, \&
  \bibinfo{author}{Lef{\`e}vre, G.} (\bibinfo{year}{2012}).
\newblock \bibinfo{title}{Towards a description of particulate fouling: From
  single particle deposition to clogging}.
\newblock {\it \bibinfo{journal}{Advances in colloid and interface science}\/},
   {\it \bibinfo{volume}{185}\/}, \bibinfo{pages}{34--76}. \URLprefix
  \url{https://doi.org/10.1016/j.cis.2012.10.001}.
\bibitem[{van Hout(2013)}]{vanhout2013spatially}
\bibinfo{author}{van Hout, R.} (\bibinfo{year}{2013}).
\newblock \bibinfo{title}{Spatially and temporally resolved measurements of
  bead resuspension and saltation in a turbulent water channel flow}.
\newblock {\it \bibinfo{journal}{Journal of Fluid Mechanics}\/},  {\it
  \bibinfo{volume}{715}\/}, \bibinfo{pages}{389--423}. \URLprefix
  \url{https://doi.org/10.1017/jfm.2012.525}.
\bibitem[{Ibrahim et~al.(2003)Ibrahim, Dunn \&
  Brach}]{ibrahim2003microparticle}
\bibinfo{author}{Ibrahim, A.}, \bibinfo{author}{Dunn, P.}, \&
  \bibinfo{author}{Brach, R.} (\bibinfo{year}{2003}).
\newblock \bibinfo{title}{Microparticle detachment from surfaces exposed to
  turbulent air flow: controlled experiments and modeling}.
\newblock {\it \bibinfo{journal}{Journal of aerosol science}\/},  {\it
  \bibinfo{volume}{34}\/}, \bibinfo{pages}{765--782}. \URLprefix
  \url{https://doi.org/10.1016/S0021-8502(03)00031-4}.
\bibitem[{Ibrahim et~al.(2004)Ibrahim, Dunn \&
  Brach}]{ibrahim2004microparticle}
\bibinfo{author}{Ibrahim, A.}, \bibinfo{author}{Dunn, P.}, \&
  \bibinfo{author}{Brach, R.} (\bibinfo{year}{2004}).
\newblock \bibinfo{title}{Microparticle detachment from surfaces exposed to
  turbulent air flow: Effects of flow and particle deposition characteristics}.
\newblock {\it \bibinfo{journal}{Journal of Aerosol Science}\/},  {\it
  \bibinfo{volume}{35}\/}, \bibinfo{pages}{805--821}. \URLprefix
  \url{https://doi.org/10.1016/j.jaerosci.2004.01.002}.
\bibitem[{Ihalainen et~al.(2014)Ihalainen, Lind, Arffman, Torvela \&
  Jokiniemi}]{ihalainen2014break}
\bibinfo{author}{Ihalainen, M.}, \bibinfo{author}{Lind, T.},
  \bibinfo{author}{Arffman, A.}, \bibinfo{author}{Torvela, T.}, \&
  \bibinfo{author}{Jokiniemi, J.} (\bibinfo{year}{2014}).
\newblock \bibinfo{title}{Break-up and bounce of tio2 agglomerates by
  impaction}.
\newblock {\it \bibinfo{journal}{Aerosol science and technology}\/},  {\it
  \bibinfo{volume}{48}\/}, \bibinfo{pages}{31--41}. \URLprefix
  \url{https://doi.org/10.1080/02786826.2013.852155}.
\bibitem[{Iimura et~al.(2009)Iimura, Watanabe, Suzuki, Hirota \&
  Higashitani}]{iimura2009simulation}
\bibinfo{author}{Iimura, K.}, \bibinfo{author}{Watanabe, S.},
  \bibinfo{author}{Suzuki, M.}, \bibinfo{author}{Hirota, M.}, \&
  \bibinfo{author}{Higashitani, K.} (\bibinfo{year}{2009}).
\newblock \bibinfo{title}{Simulation of entrainment of agglomerates from plate
  surfaces by shear flows}.
\newblock {\it \bibinfo{journal}{Chemical Engineering Science}\/},  {\it
  \bibinfo{volume}{64}\/}, \bibinfo{pages}{1455--1461}. \URLprefix
  \url{https://doi.org/10.1016/j.ces.2008.10.070}.
\bibitem[{Israelachvili(2011)}]{israelachvili2011intermolecular}
\bibinfo{author}{Israelachvili, J.~N.} (\bibinfo{year}{2011}).
\newblock {\it \bibinfo{title}{Intermolecular and surface forces}\/}.
\newblock \bibinfo{publisher}{Academic press}.
\bibitem[{Jaiswal et~al.(2009)Jaiswal, Kumar, Kilroy \&
  Beaudoin}]{jaiswal2009modeling}
\bibinfo{author}{Jaiswal, R.~P.}, \bibinfo{author}{Kumar, G.},
  \bibinfo{author}{Kilroy, C.~M.}, \& \bibinfo{author}{Beaudoin, S.~P.}
  (\bibinfo{year}{2009}).
\newblock \bibinfo{title}{Modeling and validation of the van der waals force
  during the adhesion of nanoscale objects to rough surfaces: A detailed
  description}.
\newblock {\it \bibinfo{journal}{Langmuir}\/},  {\it \bibinfo{volume}{25}\/},
  \bibinfo{pages}{10612--10623}. \URLprefix
  \url{http://pubs.acs.org/doi/abs/10.1021/la804275m}.
\bibitem[{Jim{\'e}nez(2004)}]{jimenez2004turbulent}
\bibinfo{author}{Jim{\'e}nez, J.} (\bibinfo{year}{2004}).
\newblock \bibinfo{title}{Turbulent flows over rough walls}.
\newblock {\it \bibinfo{journal}{Annu. Rev. Fluid Mech.}\/},  {\it
  \bibinfo{volume}{36}\/}, \bibinfo{pages}{173--196}. \URLprefix
  \url{https://doi.org/10.1146/annurev.fluid.36.050802.122103}.
\bibitem[{Johnson et~al.(1971)Johnson, Kendall \& Roberts}]{johnson1971surface}
\bibinfo{author}{Johnson, K.}, \bibinfo{author}{Kendall, K.}, \&
  \bibinfo{author}{Roberts, A.} (\bibinfo{year}{1971}).
\newblock \bibinfo{title}{Surface energy and the contact of elastic solids}.
\newblock In {\it \bibinfo{booktitle}{Proceedings of the Royal Society of
  London A: Mathematical, Physical and Engineering Sciences}\/} (pp.
  \bibinfo{pages}{301--313}).
\newblock \bibinfo{organization}{The Royal Society} volume
  \bibinfo{volume}{324}.
\bibitem[{Kalasin \& Santore(2015)}]{kalasin2015engineering}
\bibinfo{author}{Kalasin, S.}, \& \bibinfo{author}{Santore, M.~M.}
  (\bibinfo{year}{2015}).
\newblock \bibinfo{title}{Engineering nanoscale surface features to sustain
  microparticle rolling in flow}.
\newblock {\it \bibinfo{journal}{ACS nano}\/},  {\it \bibinfo{volume}{9}\/},
  \bibinfo{pages}{4706--4716}. \URLprefix
  \url{http://pubs.acs.org/doi/abs/10.1021/nn505322m}.
\bibitem[{Kassab et~al.(2013)Kassab, Ugaz, King \& Hassan}]{kassab2013high}
\bibinfo{author}{Kassab, A.~S.}, \bibinfo{author}{Ugaz, V.~M.},
  \bibinfo{author}{King, M.~D.}, \& \bibinfo{author}{Hassan, Y.~A.}
  (\bibinfo{year}{2013}).
\newblock \bibinfo{title}{High resolution study of micrometer particle
  detachment on different surfaces}.
\newblock {\it \bibinfo{journal}{Aerosol Science and Technology}\/},  {\it
  \bibinfo{volume}{47}\/}, \bibinfo{pages}{351--360}. \URLprefix
  \url{https://doi.org/10.1080/02786826.2012.752789}.
\bibitem[{Kim et~al.(2010)Kim, Gidwani, Wyslouzil \& Sohn}]{kim2010source}
\bibinfo{author}{Kim, Y.}, \bibinfo{author}{Gidwani, A.},
  \bibinfo{author}{Wyslouzil, B.~E.}, \& \bibinfo{author}{Sohn, C.~W.}
  (\bibinfo{year}{2010}).
\newblock \bibinfo{title}{Source term models for fine particle resuspension
  from indoor surfaces}.
\newblock {\it \bibinfo{journal}{Building and Environment}\/},  {\it
  \bibinfo{volume}{45}\/}, \bibinfo{pages}{1854--1865}. \URLprefix
  \url{https://doi.org/10.1016/j.buildenv.2010.02.016}.
\bibitem[{Kobayakawa et~al.(2015)Kobayakawa, Kiriyama, Yasuda \&
  Matsusaka}]{kobayakawa2015microscopic}
\bibinfo{author}{Kobayakawa, M.}, \bibinfo{author}{Kiriyama, S.},
  \bibinfo{author}{Yasuda, M.}, \& \bibinfo{author}{Matsusaka, S.}
  (\bibinfo{year}{2015}).
\newblock \bibinfo{title}{Microscopic analysis of particle detachment from an
  obliquely oscillating plate}.
\newblock {\it \bibinfo{journal}{Chemical Engineering Science}\/},  {\it
  \bibinfo{volume}{123}\/}, \bibinfo{pages}{388--394}. \URLprefix
  \url{https://doi.org/10.1016/j.ces.2014.11.046}.
\bibitem[{Kok et~al.(2012)Kok, Parteli, Michaels \& Karam}]{kok2012physics}
\bibinfo{author}{Kok, J.~F.}, \bibinfo{author}{Parteli, E.~J.},
  \bibinfo{author}{Michaels, T.~I.}, \& \bibinfo{author}{Karam, D.~B.}
  (\bibinfo{year}{2012}).
\newblock \bibinfo{title}{The physics of wind-blown sand and dust}.
\newblock {\it \bibinfo{journal}{Reports on Progress in Physics}\/},  {\it
  \bibinfo{volume}{75}\/}, \bibinfo{pages}{106901}. \URLprefix
  \url{https://doi.org/10.1088/0034-4885/75/10/106901}.
\bibitem[{Kubota \& Higuchi(2013)}]{kubota2013aerodynamic}
\bibinfo{author}{Kubota, Y.}, \& \bibinfo{author}{Higuchi, H.}
  (\bibinfo{year}{2013}).
\newblock \bibinfo{title}{Aerodynamic particle resuspension due to human foot
  and model foot motions}.
\newblock {\it \bibinfo{journal}{Aerosol Science and Technology}\/},  {\it
  \bibinfo{volume}{47}\/}, \bibinfo{pages}{208--217}. \URLprefix
  \url{https://doi.org/10.1080/02786826.2012.742486}.
\bibitem[{Lancaster(1994)}]{lancaster1994dune}
\bibinfo{author}{Lancaster, N.} (\bibinfo{year}{1994}).
\newblock \bibinfo{title}{Dune morphology and dynamics}.
\newblock In {\it \bibinfo{booktitle}{Geomorphology of desert environments}\/}
  (pp. \bibinfo{pages}{474--505}).
\newblock \bibinfo{publisher}{Springer}.
\bibitem[{Liang et~al.(2007)Liang, Hilal, Langston \&
  Starov}]{liang2007interaction}
\bibinfo{author}{Liang, Y.}, \bibinfo{author}{Hilal, N.},
  \bibinfo{author}{Langston, P.}, \& \bibinfo{author}{Starov, V.}
  (\bibinfo{year}{2007}).
\newblock \bibinfo{title}{Interaction forces between colloidal particles in
  liquid: Theory and experiment}.
\newblock {\it \bibinfo{journal}{Advances in colloid and interface science}\/},
   {\it \bibinfo{volume}{134}\/}, \bibinfo{pages}{151--166}. \URLprefix
  \url{https://doi.org/10.1016/j.cis.2007.04.003}.
\bibitem[{Lundell et~al.(2011)Lundell, S{\"o}derberg \&
  Alfredsson}]{lundell2011fluid}
\bibinfo{author}{Lundell, F.}, \bibinfo{author}{S{\"o}derberg, L.~D.}, \&
  \bibinfo{author}{Alfredsson, P.~H.} (\bibinfo{year}{2011}).
\newblock \bibinfo{title}{Fluid mechanics of papermaking}.
\newblock {\it \bibinfo{journal}{Annual Review of Fluid Mechanics}\/},  {\it
  \bibinfo{volume}{43}\/}, \bibinfo{pages}{195--217}. \URLprefix
  \url{https://doi.org/10.1146/annurev-fluid-122109-160700}.
\bibitem[{Marchioli et~al.(2003)Marchioli, Giusti, Salvetti \&
  Soldati}]{marchioli2003direct}
\bibinfo{author}{Marchioli, C.}, \bibinfo{author}{Giusti, A.},
  \bibinfo{author}{Salvetti, M.~V.}, \& \bibinfo{author}{Soldati, A.}
  (\bibinfo{year}{2003}).
\newblock \bibinfo{title}{Direct numerical simulation of particle wall transfer
  and deposition in upward turbulent pipe flow}.
\newblock {\it \bibinfo{journal}{International journal of Multiphase flow}\/},
  {\it \bibinfo{volume}{29}\/}, \bibinfo{pages}{1017--1038}. \URLprefix
  \url{https://doi.org/10.1016/S0301-9322(03)00036-3}.
\bibitem[{Marchioli \& Soldati(2002)}]{marchioli2002mechanisms}
\bibinfo{author}{Marchioli, C.}, \& \bibinfo{author}{Soldati, A.}
  (\bibinfo{year}{2002}).
\newblock \bibinfo{title}{Mechanisms for particle transfer and segregation in a
  turbulent boundary layer}.
\newblock {\it \bibinfo{journal}{Journal of fluid Mechanics}\/},  {\it
  \bibinfo{volume}{468}\/}, \bibinfo{pages}{283--315}. \URLprefix
  \url{https://doi.org/10.1017/S0022112002001738}.
\bibitem[{Marshall \& Li(2014)}]{marshall2014adhesive}
\bibinfo{author}{Marshall, J.~S.}, \& \bibinfo{author}{Li, S.}
  (\bibinfo{year}{2014}).
\newblock {\it \bibinfo{title}{Adhesive particle flow}\/}.
\newblock \bibinfo{publisher}{Cambridge University Press}.
\bibitem[{Matsusaka(2015)}]{matsusaka2015high}
\bibinfo{author}{Matsusaka, S.} (\bibinfo{year}{2015}).
\newblock \bibinfo{title}{High-resolution analysis of particle deposition and
  resuspension in turbulent channel flow}.
\newblock {\it \bibinfo{journal}{Aerosol Science and Technology}\/},  {\it
  \bibinfo{volume}{49}\/}, \bibinfo{pages}{739--746}. \URLprefix
  \url{https://doi.org/10.1080/02786826.2015.1066752}.
\bibitem[{McClung \& Schaerer(2006)}]{mcclung2006avalanche}
\bibinfo{author}{McClung, D.}, \& \bibinfo{author}{Schaerer, P.~A.}
  (\bibinfo{year}{2006}).
\newblock {\it \bibinfo{title}{The avalanche handbook}\/}.
\newblock \bibinfo{publisher}{The Mountaineers Books}.
\bibitem[{Minier(2016)}]{minier2016statistical}
\bibinfo{author}{Minier, J.-P.} (\bibinfo{year}{2016}).
\newblock \bibinfo{title}{Statistical descriptions of polydisperse turbulent
  two-phase flows}.
\newblock {\it \bibinfo{journal}{Physics Reports}\/},  {\it
  \bibinfo{volume}{665}\/}, \bibinfo{pages}{1--122}. \URLprefix
  \url{https://doi.org/10.1016/j.physrep.2016.10.007}.
\bibitem[{Minier \& Pozorski(2016)}]{minier2016particles}
\bibinfo{author}{Minier, J.-P.}, \& \bibinfo{author}{Pozorski, J.}
  (\bibinfo{year}{2016}).
\newblock {\it \bibinfo{title}{Particles in Wall-bounded Turbulent Flows:
  Deposition, Re-suspension and Agglomeration}\/} volume \bibinfo{volume}{571}.
\newblock \bibinfo{publisher}{Springer}.
\newblock \URLprefix \url{https://doi.org/10.1007/978-3-319-41567-3}.
\bibitem[{Mollinger \& Nieuwstadt(1996)}]{mollinger1996measurement}
\bibinfo{author}{Mollinger, A.}, \& \bibinfo{author}{Nieuwstadt, F.}
  (\bibinfo{year}{1996}).
\newblock \bibinfo{title}{Measurement of the lift force on a particle fixed to
  the wall in the viscous sublayer of a fully developed turbulent boundary
  layer}.
\newblock {\it \bibinfo{journal}{Journal of Fluid Mechanics}\/},  {\it
  \bibinfo{volume}{316}\/}, \bibinfo{pages}{285--306}. \URLprefix
  \url{https://doi.org/10.1017/S0022112096000547}.
\bibitem[{Mustin \& Stoeber(2010)}]{mustin2010deposition}
\bibinfo{author}{Mustin, B.}, \& \bibinfo{author}{Stoeber, B.}
  (\bibinfo{year}{2010}).
\newblock \bibinfo{title}{Deposition of particles from polydisperse suspensions
  in microfluidic systems}.
\newblock {\it \bibinfo{journal}{Microfluidics and nanofluidics}\/},  {\it
  \bibinfo{volume}{9}\/}, \bibinfo{pages}{905--913}. \URLprefix
  \url{https://doi.org/10.1007/s10404-010-0613-4}.
\bibitem[{Nickling \& Neuman(2009)}]{nickling2009aeolian}
\bibinfo{author}{Nickling, W.~G.}, \& \bibinfo{author}{Neuman, C.~M.}
  (\bibinfo{year}{2009}).
\newblock \bibinfo{title}{Aeolian sediment transport}.
\newblock In {\it \bibinfo{booktitle}{Geomorphology of desert environments}\/}
  (pp. \bibinfo{pages}{517--555}).
\newblock \bibinfo{publisher}{Springer}.
\bibitem[{O'neill(1968)}]{o1968sphere}
\bibinfo{author}{O'neill, M.} (\bibinfo{year}{1968}).
\newblock \bibinfo{title}{A sphere in contact with a plane wall in a slow
  linear shear flow}.
\newblock {\it \bibinfo{journal}{Chemical Engineering Science}\/},  {\it
  \bibinfo{volume}{23}\/}, \bibinfo{pages}{1293--1298}. \URLprefix
  \url{https://doi.org/10.1016/0009-2509(68)89039-6}.
\bibitem[{Persson \& Tosatti(2013)}]{persson2013physics}
\bibinfo{author}{Persson, B.}, \& \bibinfo{author}{Tosatti, E.}
  (\bibinfo{year}{2013}).
\newblock {\it \bibinfo{title}{Physics of sliding friction}\/} volume
  \bibinfo{volume}{311}.
\newblock \bibinfo{publisher}{Springer Science \& Business Media}.
\bibitem[{Persson(2013)}]{persson2013sliding}
\bibinfo{author}{Persson, B.~N.} (\bibinfo{year}{2013}).
\newblock {\it \bibinfo{title}{Sliding friction: physical principles and
  applications}\/}.
\newblock \bibinfo{publisher}{Springer Science \& Business Media}.
\bibitem[{Pope(2000)}]{pope2000turbulent}
\bibinfo{author}{Pope, S.~B.} (\bibinfo{year}{2000}).
\newblock {\it \bibinfo{title}{Turbulent flows}\/}.
\newblock \bibinfo{publisher}{Cambridge University Press}.
\bibitem[{Prokopovich \& Perni(2010)}]{prokopovich2010multiasperity}
\bibinfo{author}{Prokopovich, P.}, \& \bibinfo{author}{Perni, S.}
  (\bibinfo{year}{2010}).
\newblock \bibinfo{title}{Multiasperity contact adhesion model for universal
  asperity height and radius of curvature distributions}.
\newblock {\it \bibinfo{journal}{Langmuir}\/},  {\it \bibinfo{volume}{26}\/},
  \bibinfo{pages}{17028--17036}. \URLprefix
  \url{http://pubs.acs.org/doi/abs/10.1021/la102208y}.
\bibitem[{Prokopovich \& Starov(2011)}]{prokopovich2011adhesion}
\bibinfo{author}{Prokopovich, P.}, \& \bibinfo{author}{Starov, V.}
  (\bibinfo{year}{2011}).
\newblock \bibinfo{title}{Adhesion models: From single to multiple asperity
  contacts}.
\newblock {\it \bibinfo{journal}{Advances in colloid and interface science}\/},
   {\it \bibinfo{volume}{168}\/}, \bibinfo{pages}{210--222}. \URLprefix
  \url{https://doi.org/10.1016/j.cis.2011.03.004}.
\bibitem[{Rabinovich \& Kalman(2009{\natexlab{a}})}]{rabinovich2009incipientA}
\bibinfo{author}{Rabinovich, E.}, \& \bibinfo{author}{Kalman, H.}
  (\bibinfo{year}{2009}{\natexlab{a}}).
\newblock \bibinfo{title}{Incipient motion of individual particles in
  horizontal particle--fluid systems: A. experimental analysis}.
\newblock {\it \bibinfo{journal}{Powder Technology}\/},  {\it
  \bibinfo{volume}{192}\/}, \bibinfo{pages}{318--325}. \URLprefix
  \url{https://doi.org/10.1016/j.powtec.2009.01.013}.
\bibitem[{Rabinovich \& Kalman(2009{\natexlab{b}})}]{rabinovich2009incipientB}
\bibinfo{author}{Rabinovich, E.}, \& \bibinfo{author}{Kalman, H.}
  (\bibinfo{year}{2009}{\natexlab{b}}).
\newblock \bibinfo{title}{Incipient motion of individual particles in
  horizontal particle--fluid systems: B. theoretical analysis}.
\newblock {\it \bibinfo{journal}{Powder Technology}\/},  {\it
  \bibinfo{volume}{192}\/}, \bibinfo{pages}{326--338}. \URLprefix
  \url{https://doi.org/10.1016/j.powtec.2009.01.014}.
\bibitem[{Rabinovich et~al.(2000{\natexlab{a}})Rabinovich, Adler, Ata, Singh \&
  Moudgil}]{rabinovich2000adhesionI}
\bibinfo{author}{Rabinovich, Y.~I.}, \bibinfo{author}{Adler, J.~J.},
  \bibinfo{author}{Ata, A.}, \bibinfo{author}{Singh, R.~K.}, \&
  \bibinfo{author}{Moudgil, B.~M.} (\bibinfo{year}{2000}{\natexlab{a}}).
\newblock \bibinfo{title}{Adhesion between nanoscale rough surfaces: I. role of
  asperity geometry}.
\newblock {\it \bibinfo{journal}{Journal of Colloid and Interface Science}\/},
  {\it \bibinfo{volume}{232}\/}, \bibinfo{pages}{10--16}. \URLprefix
  \url{https://doi.org/10.1006/jcis.2000.7167}.
\bibitem[{Rabinovich et~al.(2000{\natexlab{b}})Rabinovich, Adler, Ata, Singh \&
  Moudgil}]{rabinovich2000adhesionII}
\bibinfo{author}{Rabinovich, Y.~I.}, \bibinfo{author}{Adler, J.~J.},
  \bibinfo{author}{Ata, A.}, \bibinfo{author}{Singh, R.~K.}, \&
  \bibinfo{author}{Moudgil, B.~M.} (\bibinfo{year}{2000}{\natexlab{b}}).
\newblock \bibinfo{title}{Adhesion between nanoscale rough surfaces: Ii.
  measurement and comparison with theory}.
\newblock {\it \bibinfo{journal}{Journal of Colloid and Interface Science}\/},
  {\it \bibinfo{volume}{232}\/}, \bibinfo{pages}{17--24}. \URLprefix
  \url{https://doi.org/10.1006/jcis.2000.7168}.
\bibitem[{Reeks \& Hall(2001)}]{reeks2001kinetic}
\bibinfo{author}{Reeks, M.}, \& \bibinfo{author}{Hall, D.}
  (\bibinfo{year}{2001}).
\newblock \bibinfo{title}{Kinetic models for particle resuspension in turbulent
  flows: theory and measurement}.
\newblock {\it \bibinfo{journal}{Journal of Aerosol Science}\/},  {\it
  \bibinfo{volume}{32}\/}, \bibinfo{pages}{1--31}. \URLprefix
  \url{https://doi.org/10.1016/S0021-8502(00)00063-X}.
\bibitem[{Ryde \& Matijevi{\'c}(2000)}]{ryde2000deposition}
\bibinfo{author}{Ryde, N.~P.}, \& \bibinfo{author}{Matijevi{\'c}, E.}
  (\bibinfo{year}{2000}).
\newblock \bibinfo{title}{Deposition and detachment studies of fine particles
  by the packed column technique}.
\newblock {\it \bibinfo{journal}{Colloids and Surfaces A: Physicochemical and
  Engineering Aspects}\/},  {\it \bibinfo{volume}{165}\/},
  \bibinfo{pages}{59--78}. \URLprefix
  \url{https://doi.org/10.1016/S0927-7757(99)00447-1}.
\bibitem[{Shaqfeh(2005)}]{shaqfeh2005dynamics}
\bibinfo{author}{Shaqfeh, E.~S.} (\bibinfo{year}{2005}).
\newblock \bibinfo{title}{The dynamics of single-molecule dna in flow}.
\newblock {\it \bibinfo{journal}{Journal of Non-Newtonian Fluid Mechanics}\/},
  {\it \bibinfo{volume}{130}\/}, \bibinfo{pages}{1--28}. \URLprefix
  \url{https://doi.org/10.1016/j.jnnfm.2005.05.011}.
\bibitem[{Shnapp \& Liberzon(2015)}]{shnapp2015comparative}
\bibinfo{author}{Shnapp, R.}, \& \bibinfo{author}{Liberzon, A.}
  (\bibinfo{year}{2015}).
\newblock \bibinfo{title}{A comparative study and a mechanistic picture of
  resuspension of large particles from rough and smooth surfaces in vortex-like
  fluid flows}.
\newblock {\it \bibinfo{journal}{Chemical Engineering Science}\/},  {\it
  \bibinfo{volume}{131}\/}, \bibinfo{pages}{129--137}. \URLprefix
  \url{https://doi.org/10.1016/j.ces.2015.03.048}.
\bibitem[{Soldati \& Marchioli(2009)}]{soldati2009physics}
\bibinfo{author}{Soldati, A.}, \& \bibinfo{author}{Marchioli, C.}
  (\bibinfo{year}{2009}).
\newblock \bibinfo{title}{Physics and modelling of turbulent particle
  deposition and entrainment: Review of a systematic study}.
\newblock {\it \bibinfo{journal}{International Journal of Multiphase Flow}\/},
  {\it \bibinfo{volume}{35}\/}, \bibinfo{pages}{827--839}. \URLprefix
  \url{https://doi.org/10.1016/j.ijmultiphaseflow.2009.02.016}.
\bibitem[{Stempniewicz \& Komen(2010)}]{stempniewicz2010comparison}
\bibinfo{author}{Stempniewicz, M.}, \& \bibinfo{author}{Komen, E.}
  (\bibinfo{year}{2010}).
\newblock \bibinfo{title}{Comparison of several resuspension models against
  measured data}.
\newblock {\it \bibinfo{journal}{Nuclear Engineering and Design}\/},  {\it
  \bibinfo{volume}{240}\/}, \bibinfo{pages}{1657--1670}. \URLprefix
  \url{https://doi.org/10.1016/j.nucengdes.2010.02.018}.
\bibitem[{Stevenson et~al.(2002)Stevenson, Thorpe \&
  Davidson}]{stevenson2002incipient}
\bibinfo{author}{Stevenson, P.}, \bibinfo{author}{Thorpe, R.}, \&
  \bibinfo{author}{Davidson, J.} (\bibinfo{year}{2002}).
\newblock \bibinfo{title}{Incipient motion of a small particle in the viscous
  boundary layer at a pipe wall}.
\newblock {\it \bibinfo{journal}{Chemical Engineering Science}\/},  {\it
  \bibinfo{volume}{57}\/}, \bibinfo{pages}{4505--4520}. \URLprefix
  \url{https://doi.org/10.1016/S0009-2509(02)00418-9}.
\bibitem[{Stronge(2004)}]{stronge2004impact}
\bibinfo{author}{Stronge, W.} (\bibinfo{year}{2004}).
\newblock {\it \bibinfo{title}{Impact mechanics}\/}.
\newblock \bibinfo{address}{Cambridge}: \bibinfo{publisher}{Cambridge
  University Press}.
\bibitem[{Theerachaisupakij et~al.(2003)Theerachaisupakij, Matsusaka, Akashi \&
  Masuda}]{theerachaisupakij2003reentrainment}
\bibinfo{author}{Theerachaisupakij, W.}, \bibinfo{author}{Matsusaka, S.},
  \bibinfo{author}{Akashi, Y.}, \& \bibinfo{author}{Masuda, H.}
  (\bibinfo{year}{2003}).
\newblock \bibinfo{title}{Reentrainment of deposited particles by drag and
  aerosol collision}.
\newblock {\it \bibinfo{journal}{Journal of Aerosol Science}\/},  {\it
  \bibinfo{volume}{34}\/}, \bibinfo{pages}{261--274}. \URLprefix
  \url{https://doi.org/10.1016/S0021-8502(02)00180-5}.
\bibitem[{Traugott \& Liberzon(2017)}]{traugott2017experimental}
\bibinfo{author}{Traugott, H.}, \& \bibinfo{author}{Liberzon, A.}
  (\bibinfo{year}{2017}).
\newblock \bibinfo{title}{Experimental study of forces on freely moving
  spherical particles during resuspension into turbulent flow}.
\newblock {\it \bibinfo{journal}{International Journal of Multiphase Flow}\/},
  {\it \bibinfo{volume}{88}\/}, \bibinfo{pages}{167--178}. \URLprefix
  \url{https://doi.org/10.1016/j.ijmultiphaseflow.2016.10.003}.
\bibitem[{Voth \& Soldati(2017)}]{voth2017anisotropic}
\bibinfo{author}{Voth, G.~A.}, \& \bibinfo{author}{Soldati, A.}
  (\bibinfo{year}{2017}).
\newblock \bibinfo{title}{Anisotropic particles in turbulence}.
\newblock {\it \bibinfo{journal}{Annual Review of Fluid Mechanics}\/},  {\it
  \bibinfo{volume}{49}\/}, \bibinfo{pages}{249--276}. \URLprefix
  \url{https://doi.org/10.1146/annurev-fluid-010816-060135}.
\bibitem[{Wang(1990)}]{wang1990effects}
\bibinfo{author}{Wang, H.-C.} (\bibinfo{year}{1990}).
\newblock \bibinfo{title}{Effects of inceptive motion on particle detachment
  from surfaces}.
\newblock {\it \bibinfo{journal}{Aerosol Science and Technology}\/},  {\it
  \bibinfo{volume}{13}\/}, \bibinfo{pages}{386--393}. \URLprefix
  \url{https://doi.org/10.1080/02786829008959453}.
\bibitem[{Whitehouse(2004)}]{whitehouse2004surfaces}
\bibinfo{author}{Whitehouse, D.~J.} (\bibinfo{year}{2004}).
\newblock {\it \bibinfo{title}{Surfaces and their Measurement}\/}.
\newblock \bibinfo{publisher}{Elsevier}.
\bibitem[{Wu et~al.(2017)Wu, Soligo, Marchioli, Soldati \&
  Piomelli}]{wu2017particle}
\bibinfo{author}{Wu, W.}, \bibinfo{author}{Soligo, G.},
  \bibinfo{author}{Marchioli, C.}, \bibinfo{author}{Soldati, A.}, \&
  \bibinfo{author}{Piomelli, U.} (\bibinfo{year}{2017}).
\newblock \bibinfo{title}{Particle resuspension by a periodically forced
  impinging jet}.
\newblock {\it \bibinfo{journal}{Journal of Fluid Mechanics}\/},  {\it
  \bibinfo{volume}{820}\/}, \bibinfo{pages}{284--311}. \URLprefix
  \url{https://doi.org/10.1017/jfm.2017.210}.
\bibitem[{Zhang et~al.(2013)Zhang, Reeks \& Kissane}]{zhang2013particle}
\bibinfo{author}{Zhang, F.}, \bibinfo{author}{Reeks, M.}, \&
  \bibinfo{author}{Kissane, M.} (\bibinfo{year}{2013}).
\newblock \bibinfo{title}{Particle resuspension in turbulent boundary layers
  and the influence of non-gaussian removal forces}.
\newblock {\it \bibinfo{journal}{Journal of Aerosol Science}\/},  {\it
  \bibinfo{volume}{58}\/}, \bibinfo{pages}{103--128}. \URLprefix
  \url{https://doi.org/10.1016/j.jaerosci.2012.11.009}.
\bibitem[{Zhang \& Ahmadi(2011)}]{zhang2011effects}
\bibinfo{author}{Zhang, X.}, \& \bibinfo{author}{Ahmadi, G.}
  (\bibinfo{year}{2011}).
\newblock \bibinfo{title}{Effects of electrostatic and capillary forces and
  surface deformation on particle detachment in turbulent flows}.
\newblock {\it \bibinfo{journal}{Journal of Adhesion Science and
  Technology}\/},  {\it \bibinfo{volume}{25}\/}, \bibinfo{pages}{1175--1210}.
  \URLprefix \url{https://doi.org/10.1163/016942410X549906}.
\bibitem[{Ziskind(2006)}]{ziskind2006particle}
\bibinfo{author}{Ziskind, G.} (\bibinfo{year}{2006}).
\newblock \bibinfo{title}{Particle resuspension from surfaces: Revisited and
  re-evaluated}.
\newblock {\it \bibinfo{journal}{Reviews in Chemical Engineering}\/},  {\it
  \bibinfo{volume}{22}\/}, \bibinfo{pages}{1--123}. \URLprefix
  \url{https://doi.org/10.1515/REVCE.2006.22.1-2.1}.
\bibitem[{Ziskind et~al.(1995)Ziskind, Fichman \&
  Gutfinger}]{ziskind1995resuspension}
\bibinfo{author}{Ziskind, G.}, \bibinfo{author}{Fichman, M.}, \&
  \bibinfo{author}{Gutfinger, C.} (\bibinfo{year}{1995}).
\newblock \bibinfo{title}{Resuspension of particulates from surfaces to
  turbulent flows—review and analysis}.
\newblock {\it \bibinfo{journal}{Journal of Aerosol Science}\/},  {\it
  \bibinfo{volume}{26}\/}, \bibinfo{pages}{613--644}. \URLprefix
  \url{https://doi.org/10.1016/0021-8502(94)00139-P}.

\end{thebibliography}
\bibliographystyle{model5-names}

\end{document}